\documentclass[msc, maledoc,pdftex,numbers]{coppe}
\usepackage[english,brazil]{babel}
\usepackage{hyphenat}
\usepackage[utf8]{inputenc}
\usepackage{amsmath,amssymb}
\usepackage[]{amsthm}
\usepackage{lmodern}

\usepackage{graphicx}
\usepackage{subfig}
\usepackage{float}
\usepackage{epsfig}
\usepackage{epstopdf}
\usepackage{placeins} 

\usepackage[]{multirow}
\usepackage[]{hhline}

\usepackage{caption}
\usepackage[boxed]{algorithm2e}

\SetKwInput{KwIn}{Entrada}
\SetKwInput{KwData}{Entrada}
\SetKwInput{KwResult}{Salida}
\SetKwInput{KwOut}{Salida}
\SetKwInput{Return}{retornar}
\SetKwFor{For}{desde}{hacer}{fin\_desde}
\SetKwFor{While}{mientras}{hacer}{fin\_mientras}
\SetKwIF{If}{ElseIf}{Else}{si}{entonces}{si\_no si}{si\_no}{fin\_si}

\makelosymbols
\makeloabbreviations
\graphicspath{{./structurePlot/}{./experiments/}{./bloques/}{./stagnation/}{./helmke/}}
\DeclareGraphicsExtensions{{.eps}}

\begin{document}
% \newcommand{\myalfnt}[1]{\small #1}
% \SetAlFnt{\myalfnt}
\renewcommand\contentsname{ÍNDICE GENERAL}
\renewcommand\listfigurename{LISTA DE FIGURAS}
\renewcommand\listtablename{LISTA DE TABLAS}
\renewcommand\bibname{REFERENCIAS BIBLIOGRÁFICAS}
\renewcommand\indexname{Indice alfabético}
\renewcommand\figurename{Figura}
\renewcommand\tablename{Tabla}
\renewcommand\partname{Parte}
\renewcommand\chaptername{Capítulo}
\renewcommand\appendixname{APENDICE}
\renewcommand\abstractname{RESUMEN}
\renewcommand\proofname{Demostración}
\SetAlgorithmName{Algoritmo}{algoritmo}{LISTA DE ALGORITMOS} 

  \title{REPRESENTATIVIDAD MUESTRAL EN LA INCERTIDUMBRE SIMÉTRICA MULTIVARIADA PARA LA SELECCIÓN DE ATRIBUTOS}
  \foreigntitle{SAMPLE REPRESENTATIVENESS IN MULTIVARIATE SYMMETRICAL UNCERTAINTY FOR FEATURE SELECTION}
  %\author{*}{*}
  %\advisor{*}{*}{*}
  \author{Gustavo Daniel}{Sosa Cabrera}
  \advisor{Prof.}{Miguel}{Garc\'ia Torres}{D.Sc.}
  \advisor{Prof.}{Santiago}{G\'omez}{M.Sc.}
  \advisor{Prof.}{Christian E.}{Schaerer Serra}{D.Sc.}

  \examiner{Prof.}{Daniel Romero}{Dr.}
  \examiner{Prof.}{Cynthia Villalba}{Dra.}
  \examiner{Prof.}{Sebastián Grillo}{D.Sc.}
  \examiner{Prof.}{Diego Stalder}{D.Sc.}
  \examiner{Prof.}{Santiago Gómez}{M.Sc.}
  \examiner{Prof.}{Christian Schaerer}{D.Sc.}

  \university{Universidad Nacional de Asunci\'on}
  \place{Asunci\'on}{}{Paraguay}

  \department{PEC}
  \date{7}{2017}
  \maketitle

%-- 
% gdsosa:
% Error de compilacion, se modifico la linea 333 de coppe.cls
% Ver: https://sourceforge.net/p/coppetex/mailman/coppetex-users/ 
  %\frontmatter
%--
  \dedication{Dedicado a mi amada esposa Leslie, a mis adorados padres Epifanio y Teresa, y a mi apreciado hermano Osvaldo.}
  \chapter*{Agradecimientos}

Al Laboratorio de Computaci\'on Cient\'ifica y Aplicada (LCCA) de la Facultad Politécnica de la Universidad Nacional de Asunci\'on (FP-UNA).

%\noindent A la Facultad Politécnica por facilitar mi viaje a la Universidad Federal de Rio de Janeiro (UFRJ) para trabajar con el Prof. Bhaya.

%\noindent Al NACAD (Nucleo de Computación de Alto Desempeño) de la UFRJ, por haberme recibido durante mi estancia en Rio de Janeiro.

\noindent A mis orientadores: los profesores D.Sc. Miguel García Torres, M.Sc. Santiago Gómez Guerrero y D.Sc. Christian Schaerer Serra, con su gu\'ia este trabajo ha finalizado exitosamente.

  \begin{abstract}
\noindent En el presente trabajo hemos analizado el comportamiento de una versión multivariada de la incertidumbre simétrica a través de técnicas de simulación estadísticas sobre varias combinaciones de atributos informativos y no-informativos generados de forma aleatoria. Los experimentos muestran como el número de atributos, sus cardinalidades y el tamaño muestral afectan al MSU como medida. En esta tesis, mediante la observación de resultados hemos propuesto una condición que preserva una buena calidad en el MSU bajo diferentes combinaciones de los tres factores mencionados, lo cual provee un nuevo y valioso criterio para llevar a cabo el proceso de reducción de dimensionalidad.

\vspace{0.1in}
 \noindent {\bf Palabras claves}: {\it selección de atributos, incertidumbre simétrica, predicción multivariada de la respuesta, alta dimensionalidad.}

\end{abstract}

  \begin{foreignabstract}

\noindent In this work, we analyze the behavior of the multivariate symmetric uncertainty (MSU) measure through the use of 
statistical simulation techniques under various mixes of informative 
and non-informative randomly generated features.
Experiments show how the number of attributes, their cardinalities, 
and the sample size affect the MSU. In this thesis, through observation 
of results, it is proposed an heuristic condition 
that preserves good quality in the MSU under different 
combinations of these three factors, providing a new useful 
criterion to help drive the process of 
dimension reduction.

\vspace{0.1in}
 \noindent {\bf Keywords}: {\it feature selection, symmetrical uncertainty, 
multivariate prediction of response, high dimensionality.}

\end{foreignabstract}

  \tableofcontents
  \listoffigures
  \listoftables 
  \printlosymbols
  \printloabbreviations

  \mainmatter
  \chapter{INTRODUCCIÓN}

El gran avance de la tecnología acaecido en los últimos años en lo referente a la recolección y almacenamiento de datos, ha desencadenado en la proliferación de conjuntos de datos complejos, heterogéneos y de gran envergadura, lo cual representa un reto cada vez mayor para la extracción de conocimiento; un conocimiento el cual se espera sea valioso en la toma de decisiones en el ámbito correspondiente.

%las tareas de clasificación en general y la selección de atributos en particular.

En \cite{fayyad_et_al} el descubrimiento de conocimiento en bases de datos (del inglés \textit{Knowledge Discovery in Databases}, KDD) se define como ``un proceso no trivial de identificación de patrones válidos, novedosos, parcialmente útiles y, en última instancia, comprensibles a partir de los datos".

El $KDD$ es un proceso que consta de una secuencia iterativa de etapas donde la fase de minería de datos es la más característica a tal punto que muchas veces dicha fase es utilizada para nombrar todo el proceso.

En \cite{witten} se define la minería de datos como ``el proceso de extraer conocimiento útil y comprensible desde grandes cantidades de datos almacenados en distintos formatos”. Por tanto, el objetivo principal es encontrar modelos
inteligibles a partir de estos datos.

Dentro de la minería de datos hemos de distinguir tipos de tareas las cuales pueden ser predictivas o descriptivas. Entre las tareas predictivas encontramos la clasificación y la regresión, mientras que el agrupamiento y las reglas de asociación son tareas descriptivas.

En la tarea de clasificación cada instancia pertenece a una clase la cual se indica mediante el valor discreto de un atributo que llamamos la \textit{clase} de la instancia. Por tanto, el objetivo es maximizar el poder predictivo de la clasificación de nuevas instancias de las que se desconoce la clase del caso.  

Cabe destacar, que es ampliamente aceptado, el hecho que de acuerdo con la información sobre el concepto objetivo (la clase del caso), los atributos pueden ser clasificados como relevantes, irrelevantes y las que son redundantes.

Un atributo es considerado irrelevante si el mismo no contiene información acerca de la clase y por tanto no es necesario en absoluto para la tarea de predicción. Por el contrario, un atributo es relevante si el mismo contiene información acerca de la clase y finalmente, un atributo se define como redundante si el mismo contiene información acerca de la clase que ya se encuentra disponible ya sea en otro atributo o en un subconjunto de los mismos.

Por otra parte, actualmente existen campos como por ejemplo el procesamiento de documentos y la bio-informática en los cuales sus conjuntos de datos (\textit{dataset}) son caracterizados por poseer una gran cantidad de atributos (variables) respecto a la cantidad de instancias (ejemplos).

En estos espacios, denominados de alta dimensionalidad, la selección de atributos es un método para excluir aquellas características que no son informativas (irrelevantes), como así también aquellas que poseen información parcial o totalmente redundante, cuya presencia puede complicar la tarea de clasificación.

Por tanto, la selección de atributos consiste en encontrar el mínimo conjunto de características relevantes con el objetivo de minimizar el error en el proceso de clasificación. Esto es, mediante una evaluación de atributos.

En este sentido, se han propuesto varios criterios para evaluar los atributos y determinar su importancia. Cabe destacar que basado en los criterios de evaluación, los métodos de selección de atributos pueden ser divididos en envolvente (\textit{wrapper}), filtro (\textit{filter}) y embebido (\textit{embedded}).

En los métodos del tipo envolvente, la evaluación del subconjunto de atributos encontrados durante el proceso de búsqueda se realiza por medio del propio algoritmo de aprendizaje como una especie de caja negra. 

En consecuencia, esta estrategia posee una alta precisión en cuanto a la calidad de los conjuntos de atributos, sin embargo, es costoso en términos de recursos computacionales como así también presenta un alto riesgo de sobre-ajuste, el cual es el efecto de sobre-entrenar un algoritmo de aprendizaje haciendo que el modelo se ajuste muy bien a los datos existentes pero tenga un pobre rendimiento para predecir nuevos resultados.

En los métodos del tipo filtro, la evaluación del subconjunto de atributos se lleva a cabo por medio de la valoración de las propiedades intrínsecas del dato, tales como la distancia, la consistencia, la entropía y la correlación. 

Esta estrategia no considera ninguna interacción con el algoritmo de aprendizaje por lo que son mucho más eficientes en términos de recursos computacionales que los métodos tipo envolvente, pero en contrapartida los resultados de la clasificación podrían verse desfavorecidos ya que puede considerarse como una etapa previa e independiente para el filtrado de atributos.

Finalmente, en los métodos tipo embebido, la selección de atributos está incluida en el mismo como una parte no separable, es decir, se realiza la selección de atributos durante la propia inducción del clasificador.

En el enfoque de tipo filtro, las medidas basadas en entropía han ganado notoriedad para la identificación de los atributos como informativos (relevantes) o no-informativos (no relevantes). 

Por ejemplo, en \cite{huan} se ha propuesto la utilización de la medida de correlación no-lineal denominada Incertidumbre Simétrica (SU, \textit{Symmetrical Uncertainty}) \cite{fayyad}, el cual es una medida basada en la teoría de la información que se define como el producto de una normalización de la Ganancia de la Información (IG, \textit{Information Gain})\cite{quinlan} con respecto a la entropía.

Atendiendo que podemos referirnos como variables indistintamente a los atributos o la clase, cabe destacar que el $SU$ es una medida bi-variada ya que mide la reducción de la incertidumbre (entropía) de una variable $X$, debido al conocimiento del valor de otra variable $Y$. Es decir, determina la correlación o dependencia mutua entre las dos variables. 

A pesar de los buenos resultados obtenidos con el empleo del $SU$, por su naturaleza bi-variada no es posible utilizar esta medida en los casos donde se requiera establecer la relación entre más de dos variables.

Consecuentemente, con el objeto de considerar posibles interacciones que pueda existir entre más de dos variables, la denominada Incertidumbre Simétrica Multivariada (MSU, \textit{Multivariate Symmetrical Uncertainty}) fue propuesta en \cite{Arias-Michel}, como una extensión o generalización de la medida bi-variada.

Esto es importante ya que de forma heurística se verifica que en ciertos casos, el concepto objetivo es explicado mediante la interacción de dos o más atributos. En tales casos, estos atributos serían considerados irrelevantes puesto que de manera individual parecen no aportar información acerca de la clase.

Cabe mencionar que los valores de las variables indican categorías distintivas sin implicar un orden específico, es decir, que las palabras, letras o símbolos utilizados como valores posibles son simplemente etiquetas; nos referiremos inicialmente al número de etiquetas como la \textit{cardinalidad} de la variable.  

Por otra parte, un problema general en la selección de atributos que utilizan las medidas basadas en la teoría de la información, radica en que dichas medidas poseen un sesgo en favor de aquellos atributos con mayor cardinalidad como se demuestra en \cite{liu}. Es decir, estos tipos de atributos son sistemáticamente sobrestimados con mayor valor informativo como resultado de la evaluación.

En el caso del $MSU$, al ser una extensión del $SU$ y ésta a su vez una normalización de la $IG$, por transitividad el mismo adolece de sesgo. 

Además de la influencia de la cardinalidad en el sesgo mencionado, la naturaleza multivariada del $MSU$ hace que su sesgo pueda verse afectado por la cantidad de atributos (variables) a ser evaluados en función a la cantidad de instancias (casos o ejemplos de muestra) con el que se dispone y al cual nos referiremos como \textit{tamaño muestral}.

En un muestreo es importante que todos los miembros de la población se encuentren adecuadamente representados en la muestra. En el caso que no se produzca esta adecuada representación, se produce un sesgo en el resultado. Este tipo de sesgo se denomina subcobertura, denotando que el subconjunto de la población tomado como muestra no consigue cubrir todo el espectro posible. 

Aunque el tamaño y la cobertura de la muestra de hecho son dos problemas diferentes, los mismos se encuentran fuertemente relacionados ya que ante un tamaño muestral insuficiente aumenta la probabilidad de sesgo de cobertura.

Por tanto, para evitar el sesgo de cobertura, se requiere garantizar una \textit{total representatividad} en la muestra. Este es uno de los conceptos examinados como factor explicativo del comportamiento del $MSU$. 

Dados los resultados obtenidos de los factores analizados, en el presente trabajo proponemos una relación de asociación empírica entre los mismos que permite un sesgo controlado en la evaluación de la interacción conjunta de atributos mediante el $MSU$. Esto genera la posibilidad de su aplicación en problemas caracterizados con espacios de alta dimensionalidad donde se requiera precisión en la detección de la interacción multivariada.

%Utilizamos este concepto para proponer una expresión empírica que dado un número fijo de características, ésta relaciona el tamaño muestral, el tamaño del subconjunto de características incluyendo a la clase, como así también el número de etiquetas posibles de los elementos componentes del subconjunto.

%Esta expresión nos permite establecer la gama de valores de estos factores para los cuales el $MSU$ tiene un sesgo controlado, lo cual a su vez, abre múltiples aplicaciones mediante la detección con mayor precisión de la interacción de variables entre sí como un todo, tal como para el proceso de selección de características.

%Finalmente, en base a la expresión propuesta que permite el control del sesgo en el $MSU$ por medio de un tamaño muestral que garantiza una total representatividad en la muestra, hemos llevado a cabo la simulación de pruebas de bondad de ajuste con el fin de obtener una aproximación de referencia desde la perspectiva de la inferencia estadística y que nos ha permitido llevar a cabo un análisis comparativo.

\section{Relevancia y Originalidad}
% Relevancia
La presente tesis aborda la cuestión del sesgo presente en el $MSU$ como medida de correlación conjunta de variables, esto es importante ya que sin las debidas precauciones con los factores identificados como incidentes en el comportamiento del $MSU$, los resultados estarán sesgados. Esto significa en términos coloquiales que el resultado encontrado presentará sistemáticamente un error con relación al resultado deseado. La importancia del análisis del sesgo consiste justamente en cuantificar la medida de este error. Esta tesis trata esta cuestión y presenta reglas para sobrellevarla.   

%en el tamaño muestral así como en la representatividad de los valores posibles en los atributos no son adecuadamente considerados, el resultado final presentara un sesgo con relación al resultado real buscado. Esto significa en términos coloquiales que el resultado encontrado presentara un error con relación al resultado deseado. La importancia del sesgo consiste justamente en cuantificar la medida de este error. Esta tesis analiza esta cuestión y presenta resultados para sobrellevarla.   

% Originalidad
Han sido previamente realizados varios trabajos relacionados a las medidas basadas en la teoría de la información como el $IG$ y el $SU$. Hasta el momento, si bien el sesgo es una cuestión conocida en otros ámbitos, este problema no ha sido adecuadamente tratado en el contexto de los recaudos necesarios para la selección de atributos y en especial en el uso de medidas multivariadas como el $MSU$. Esta tesis versa sobre el tratamiento del sesgo en ambos contextos mencionados. 

% Objetivo general.
El objetivo de esta tesis consiste en caracterizar el comportamiento del $MSU$, cuyos mecanismos aún no son bien conocidos. Para ello se requiere una amplia evaluación experimental orientada a establecer tanto los factores implicados como así también el posible efecto de los mismos sobre el $MSU$ como medida de correlación de variables conjuntas.

% Objetivos específicos.
Por tanto, el planteamiento de los objetivos específicos se presenta de la siguiente manera:
\begin{itemize}
\item Establecer un escenario para el estudio experimental de una medida multivariada como lo es el $MSU$.
\item Caracterizar el comportamiento del $MSU$ como una medida de la interacción conjunta de variables en el contexto de la selección de atributos.
\item Determinar los factores implicados en la precisión de los resultados consecuentes de la evaluación de la interacción conjunta de atributos mediante el $MSU$.
\item Encontrar y proponer reglas que garanticen una precisión tolerable a priori en los resultados del $MSU$ como medida de evaluación multivariada en el proceso de selección de atributos. 
\end{itemize}

% Contribuciones
Finalmente, las contribuciones del presente trabajo, se pueden resumir a grandes rasgos como sigue:
\begin{itemize}
\item Si bien existen varios trabajos que analizan el sesgo en las medidas basadas en la teoría de la información \cite{liu,kononenko,hall}, hasta el momento no se ha realizado un estudio sobre una medida multivariada como lo es el $MSU$.
\item Se determina los factores implicados en el comportamiento del $MSU$ como medida de correlación de variables agrupadas.
\item Se introduce el concepto de \textit{Total Representatividad} de la muestra como principio explicativo del comportamiento del $MSU$.
%\item Se analiza el concepto de total representatividad como una descomposición de factores y se demuestra el impacto sobre el $MSU$ de cada uno de los mismos.
\item Se propone una relación de asociación empírica entre los factores determinados que permite un comportamiento controlado del $MSU$ y que a su vez proporciona criterios para la conformación de subconjuntos a ser evaluados por el $MSU$ como parte del proceso de selección de atributos.
\item Se añade certeza a la relación de asociación empírica propuesta desde el punto de vista de la inferencia estadística, mediante la simulación de pruebas de bondad de ajuste.
\end{itemize}

\section{Organización del trabajo}
En el Capítulo 2 presentamos los fundamentos teóricos utilizados en el presente trabajo. Para ello se lleva a cabo una revisión de las nociones de la teoría de la información y de la selección de atributos. En el Capítulo 3, presentamos el análisis de sesgo realizado junto con la propuesta del presente trabajo. Para ello se describen tanto el propósito de los experimentos, como así también el escenario experimental montado y los resultados obtenidos. En el Capítulo 4 presentamos los resultados experimentales obtenidos en base a la aproximación propuesta en la presente tesis. Para ello se lleva a cabo una descripción comparativa de resultados que demuestran el logro de un comportamiento controlado del sesgo en el $MSU$ para la evaluación conjunta de atributos. Finalmente, presentamos las conclusiones y trabajos futuros en el Capítulo 5.
  \newtheorem{thm}{Teorema}%[section]
\newtheorem{lema}[thm]{Lema}%[section]
\newtheorem*{lema*}{Lema}%[section]
\newtheorem{theorem}{Teorema}[section]
\newtheorem{lemma}[theorem]{Lemma}
\newtheorem{proposition}[theorem]{Proposition}
\newtheorem{corollary}[theorem]{Corolario}
\newtheorem{definition}{Definición}
\newtheorem*{definition*}{Definición}
\SetKwInput{Kw}{Entrada} 
\chapter{FUNDAMENTOS TEÓRICOS}

En este capítulo llevaremos a cabo una revisión de las nociones de la teoría de la información y de la selección de atributos que son mencionados a lo largo del presente trabajo y cuyo interés radica en que pueden ser utilizados con el objeto de medir la cantidad de información como una reducción de la incertidumbre.

\section{Teoría de la Información}

\subsection{Entropía.}
La entropía de Shannon ($H$) \cite{shannon} de una variable aleatoria discreta $X$, con $\{x_1,\ldots,x_n\}$ como posibles valores y la función de probabilidad $P(X)$, es una medida de la incertidumbre en la predicción del siguiente valor de $X$.
\begin{definition} La entropía $H(X)$ se define como
         \begin{equation}
             H(X) := -\sum_{i} P(x_i)\log_{2}(P(x_i)),
         \end{equation}
\noindent donde $H(X)$ puede ser también interpretado como una medida de la variedad inherente a $X$, o la cantidad de información que es necesaria para predecir o describir el resultado de $X$, $P(x_i)$ es la probabilidad de la variable $X$ y la sumatoria se produce sobre todos los valores posibles de $X$, denotado por $x_i$.
\end{definition}

\noindent\textbf{Entropía Conjunta}. Para variables independientes $(X,Y)$ con $P(X,Y)$ como distribución conjunta de probabilidad se tiene la entropía conjunta $H(X,Y)$. 
\begin{definition} 
La entropía conjunta $H(X,Y)$ está definida como
\begin{equation}
	H(X,Y) := - \sum_{x \in X}
    \sum_{y \in Y} \\ P(x,y) 
    \log_2[P(x,y)].
\end{equation}
\end{definition}

\noindent\textbf{Entropía Condicional}. La entropía condicional $H(X|Y)$ cuantifica la cantidad de información necesaria para describir el resultado de $X$ dado que el valor de otra variable aleatoria discreta $Y$ es conocido.
\begin{definition} La entropía condicional está definida como
\begin{equation}
	H(X|Y) := -\sum_{j} \left[ P(y_j)\sum_{i}P(xi|y_j) 
	\log_{2}(P(x_i|y_j)) \right], 
	\nonumber
\end{equation} 
\noindent donde $P(y_j)$ es la probabilidad a \textit{priori} del valor $y_j$ de $Y$, y $P(x_i|y_j)$ es la probabilidad a \textit{posteriori} de un valor  $x_i$ para la variable $X$ puesto que el valor de la variable $Y$ es $y_j$. 
\end{definition}

\subsubsection{Propiedades de la Entropía}
\begin{enumerate}
\item $H(X) \geq 0.$
\item $H_b(X) = (\log_b a)H_a(X).$
\item $H(X|Y) \leq H(X)$ y es igual si y solamente si $X$ e $Y$ son independientes. 
\item $H(X_1,\dots,X_n) \leq \sum_{i=1}^{n} H(X_i)$ con igualdad si y solamente sí las variables aleatorias $X_i$ son independientes.
\item $H(X) \leq \log (X)$, y la igualdad se cumple si y solamente si $X$ está uniformemente distribuida en $X$.
\end{enumerate}

\begin{theorem} \label{def:regla_cadena}
(regla de la cadena) \cite{Cover}. Dadas dos variables aleatorias $X$ e $Y$ la entropía conjunta está dada por
$H(X,Y) = H(X) + H(Y|X).$
\end{theorem}

Una extensión del Teorema \ref{def:regla_cadena} para $X$ e $Y$ dado $Z$ está expresado por el corolario
\begin{corollary}
$H(X,Y|Z) = H(X|Z) + H(Y|X,Z).$
\end{corollary}

Una generalización del Teorema \ref{def:regla_cadena} está dado por
\begin{theorem}
(regla de la cadena general) \cite{Cover}. 
$H(X_1,\dots,X_n) = \sum_{i=1}^{n} H(X_i|X_1,\dots,X_{i-1}).$
\end{theorem}

\subsection{Ganancia de la Información}
Conocido alternativamente como Información Mutua \cite{SW49}, la Ganancia de la Información $(IG(X|Y))$ \cite{quinlan} de una variable $X$ con respecto a otra variable dada $Y$ mide la reducción de la incertidumbre acerca del valor de la variable $X$ cuando el valor de $Y$ es conocido.
\begin{definition} 
La información ganada de $X$ al conocer $Y$ se define como 
\begin{equation}
IG(X|Y) := H(X) - H(X|Y). 
\end{equation}
\end{definition}

$IG$ mide cuánto el conocimiento de $Y$ hace que el valor de $X$ sea más fácil de predecir, y por tanto, el mismo puede ser utilizado como una {\it medida de correlación}. 

Observe que como casos extremos son obtenidos
\begin{enumerate} %[label=(\alph*)]
\item si $X$ e $Y$ son independientes, entonces $IG(X|Y) = 0$, y 
\item si $X$ e $Y$ son completamente correlacionados entonces $H(X|Y) = 0$ 
y por tanto $IG(X|Y) = H(X)$. 
\end{enumerate}

En general, para variables aleatorias cualesquiera $X$, $Z$ e $Y$, $IG(X|Y) > IG(Z|Y)$ significa que conociendo el valor de $Y$ la reducción en la incertidumbre acerca de $X$ es mayor que la reducción en la incertidumbre acerca de $Z$, debido a que $X$ está más correlacionado a $Y$ que este a $Z$.

Se puede demostrar que $IG(X|Y)$ es una medida simétrica, lo cual es una conveniente propiedad para una medida entre dos variables ya que el orden entre ellas no altera el resultado de la medición. 

Sin embargo, $IG$ presenta un inconveniente: cuando $X$ y/o $Y$ tiene más valores posibles, ellas aparecen con mayor correlación, por tanto, $IG$ tiende a ser más alto cuando se presentan variables con mayor número de valores posibles, es decir, cuando una o ambas poseen una alta cardinalidad. La definición de cardinalidad será presentada en la Sección \ref{card_uni} más adelante.

\subsubsection{Propiedades de la ganancia de la información}
\begin{enumerate}
\item $IG(X;Y) \geq 0.$
\item $IG(X;Y) = H(X) - H(X|Y).$
\item $IG(X;Y) = H(Y) - H(Y|X).$
\item $IG(X;Y) = H(X) + H(Y) - H(X,Y).$
\item $IG(X;Y) = IG(Y;X)$ (simetría).
\item $IG(X;X) = H(X)$ (información propia).
\end{enumerate}

Dada la relación con la entropía, para la ganancia de la información se puede establecer el teorema
\begin{theorem}
(regla de la cadena) \cite{Cover}. Para un conjunto de n variables $\left\lbrace X_1,\dots,X_n \right\rbrace$ e $Y$ la ganancia de la información está dada por
$I(X_1,\dots,X_n;Y) = \sum_{i=1}^{n} I(X_i;Y|X_1,\dots,X_{i-1}).$
\end{theorem}

La unidad de información del $IG$ depende de la base del logaritmo utilizado. En el presente trabajo se ha utilizado la base 2 por lo que la unidad se encuentra en bits.
Note que el $IG$ es una \textit{semi-métrica} \cite{Kraskov} que cumple con los axiomas
\begin{enumerate}
\item $IG(X;Y) >= 0$ (no negatividad).
\item $IG(X;Y) = 0 \iff X = Y$ (identidad de los in-discernibles).
\item $IG(X;Y) = IG(Y;X)$ (simetría).
\end{enumerate}

%\subsection{Incertidumbre Simétrica}
El valor del $IG$ puede ser normalizado utilizando ambas entropías, originando la medida de Incertidumbre Simétrica ($SU$)\cite{fayyad}.
\begin{definition} 
La incertidumbre simétrica de dos variables aleatorias $X$, $Y$ se define como
\begin{equation}
	SU(X,Y) := 2 \left[ \frac{IG(X|Y)}{H(X) + H(Y)}\right].
\end{equation}
\end{definition}

Observe que, (a) si $X$ e $Y$ son independientes entonces $SU(X,Y) = 0$; y (b) si $X$ e $Y$ están completamente correlacionados entonces $IG(X|Y) = H(X) = H(Y)$ por tanto $SU(X,Y) = 1$. Como podemos apreciar, el $SU$ restringe sus valores al rango entre $0$ y $1$, es decir, $SU \in [0,1]$. 

%\subsection{Correlación Total}
De manera a generalizar la ganancia de información, se introduce la Correlación Total o Multiinformación \cite{wj,watanabe}. Esto permite establecer el nivel de correlación de $n$ variables aleatorias que conforman un conjunto.
\begin{definition}
Dado un conjunto de n variables aleatorias $\left\lbrace X_1,\dots,X_n \right\rbrace$, la correlación total se define como
\begin{equation}\label{totalcorr}
	C(X_{1:n}) := \sum_{i=1}^{n} H(X_i) - H(X_{1:n}),
\end{equation}
\noindent donde
\begin{equation}
	H(X_{1:n}) := H(X_1,\dots,X_n) := - \sum_{x_1} \dots 
    \sum_{x_n} \\ P(x_1,\dots,x_n) 
    \log_2[P(x_1,\dots,x_n)]
\end{equation}
es la entropía conjunta de las variables aleatorias $X_1,\dots, X_n$.  
\end{definition} 

%\subsection{Incertidumbre Simétrica Multivariada}
La incertidumbre simétrica multivariada ($MSU$) es propuesta en \cite{Arias-Michel} como una generalización del $SU$ basada en la correlación total (\ref{totalcorr}), con el objeto de cuantificar la redundancia o dependencia existente entre dos o más variables. 

\begin{definition} 
Dado un conjunto de n variables aleatorias $\left\lbrace X_1,\dots,X_n \right\rbrace$, la incertidumbre simétrica multivariada se define como
\begin{equation}
	MSU(X_{1:n}) := \frac{n}{n-1} \left[ \frac{C(X_{1:n})}
    {\sum_{i=1}^{n} H(X_i)}\right].
\end{equation}
\end{definition} 

Al igual que en el $SU$, el rango de valores del $MSU$ también se encuentra entre $[0,1]$. 

Cabe destacar que a diferencia de la desviación estándar y otras medidas que están orientadas a datos numéricos, la $SU$ y la $MSU$ pueden ser aplicadas a  números discretos y a variables aleatorias categóricas. Esta propiedad es conveniente para el mundo multivariado de atributos de diferentes tipos que aparecen frecuentemente en el mismo conjunto de datos. 

Además, dado que $SU$ y $MSU$ dependen únicamente de probabilidades, las mismas son invariantes ante traslaciones y cambios de escala aplicados a cualquier $X_i$, siempre y cuando la función de probabilidad permanezca igual.

\section{Selección de Atributos}

De ahora en adelante, llamaremos atributo a una variable aleatoria discreta o categórica presente en el conjunto de datos considerado.

La selección de atributos es la tarea de obtener un subconjunto de atributos del menor tamaño posible a partir de los originales presentes en un conjunto de datos dado y que proporcionen la mayor parte de la información útil. Esto es, sin que se vea afectada la predictibilidad de la clase. 

Para ello se dejan de lado los atributos detectados como \textit{irrelevantes} como también los que son \textit{redundantes}, efectuando así una reducción de dimensionalidad que resulta en un subconjunto más simple y más adecuado para la predicción deseada.

Además de la simplificación del modelo, la reducción de dimensionalidad lograda a través de una apropiada selección de atributos, disminuye el riesgo de sobreajuste (\textit{overfitting} del modelo. Un modelo sobreajustado es aquel que ha exagerado su adecuación a los casos de aprendizaje, con el consecuente deterioro en la precisión de sus predicciones para casos nuevos.
\\\\
\textbf{Relevancia de atributos}. Sea $A$ el conjunto de todos los atributos, $A_i\in A$ uno de ellos y $S_i= A-\left\lbrace A_i \right\rbrace$. Sea $C$ la clase cuya información se desea predecir.
A continuación se dará una clasificación de los atributos conforme su relevancia o irrelevancia.

\begin{definition}[Relevancia fuerte]\label{def:relevancia:fuerte}
	El atributo $A_i$ tiene una relevancia fuerte si y solamente si
	\begin{equation}
		P(C|A_i,S_i) \neq P(C|S_i),
	\end{equation}
\end{definition}
\noindent donde $P(C|A_i,S_i)$ es la probabilidad de suponer el valor de $C$ conociendo previamente los valores de $A_i$ y $S_i$.

Un atributo con relevancia fuerte es imprescindible para la solución final u óptima, es decir, no puede descartarse del conjunto $A$ sin alterar la predictibilidad de $C$.

\begin{definition}[Relevancia débil]\label{def:relevancia:debil}
	$A_i$ tiene una relevancia débil, si y solamente si,
	\begin{equation}
		P(C|A_i,S_i) = P(C|S_i)
	\end{equation}
	y existe $S'_i\subset S_i$ tal que
	\begin{equation}
		P(C|F_i,S'_i) \neq P(C|S'_i).
	\end{equation}
\end{definition}

Un atributo con relevancia débil no es relevante para $A$, es decir, no cambia la información sobre $C$ si se la descarta de $A$.
Sin embargo, sí es relevante para un subconjunto de $A$.

\begin{definition}[Irrelevancia]\label{def:irrelevancia}
	$A_i$ se considera irrelevante, si y solamente si,
	\begin{equation}
		P(C|A_i,S'_i) = P(C|S'_i) \quad \forall S'_i \subseteq S_i.
	\end{equation}
\end{definition}

Un atributo irrelevante no aporta información alguna sobre $C$, así debe descartarse para la solución final.
\\\\
\textbf{Redundancia de atributos}. Se dice que un atributo es redundante si su información está completamente contenida en uno o más atributos.

\begin{definition}[Manta de Markov]\label{def:manta:Markov}
	Dado un atributo $A_i$, sea $M_i\subset A- \left\lbrace A_i \right\rbrace$. Se dice que $M_i$ es una manta de Markov para $A_i$ si y solamente si
	\begin{equation}
		P(A-M_i-\left\lbrace A_i \right\rbrace |A,M_i) = P(A-M_i- \left\lbrace A_i \right\rbrace |M_i).
	\end{equation}
\end{definition}

Una manta de Markov de un atributo es en esencia un conjunto de atributos que contiene toda la información del atributo.
Puede verse intuitivamente como una manta que envuelve, oculta al atributo, pues ya aporta toda la información que podría aportar el atributo de los demás elementos de $A$.

\begin{definition}[Redundancia]\label{def:redundancia}
	Sea $G\subseteq A$ el conjunto actual de atributos.
	Un atributo se considera redundante (y debe ser descartado de $G$), si y solamente si, es débilmente relevante y tiene una manta de Markov dentro de $G$.
\end{definition}

De este modo, se considera un atributo redundante si su información está completamente contenida en un subconjunto de $A$ que no lo incluye.

\section{Cardinalidad Univariada y Multivariada}

La medida de cardinalidad puede ser usada con el propósito de especificar la cantidad de valores diferentes que puede tomar un objeto determinado.

En este sentido, para la selección de atributos la cardinalidad puede ser considerada con respecto a un único atributo (univariada) o con respecto a varios atributos incluyendo la clase (multivariada). 

Por tanto, para formalizar este concepto, introducimos las siguientes definiciones de cardinalidad.

\begin{definition} \label{card_uni}
Dado un atributo discreto o categórico $A$, su Cardinalidad Univariada, denotada por $\left\vert{A}\right\vert$, es el número posible de valores diferentes que puede tomar $A$.
\end{definition}

\begin{definition} \label{card_multi}
Dado un conjunto de atributos discretos o categóricos $A_1$, $A_2$,\dots, $A_n$ y $C$, donde $C$ es la clase, su Cardinalidad Multivariada es el número de posibles combinaciones de valores entre todos los atributos, incluyendo a la clase.
\end{definition}

La Definición \ref{card_uni} nos dice cuán diversa es una variable en función a los valores posibles que pueda asumir. Por otra parte, la Definición \ref{card_multi} establece la cantidad de posibles combinaciones de valores, midiendo de esta manera la diversidad de información en todo el conjunto.

\section{Sesgo de Evaluación de Atributos}

Se dice que los resultados de la evaluación de atributos están sesgados cuando existe un error sistemático en la estimación del nivel de información que poseen los atributos (individualmente o en conjunción) acerca de la clase.

En el caso de las medidas basadas en la teoría de la información, estas medidas poseen un sesgo en favor de aquellos atributos con mayor cardinalidad univariada. Es decir, estos tipos de atributos son sistemáticamente sobrestimados.

%En la selección de atributos, nos referimos como sesgo de una medida de evaluación a la diferencia entre su esperanza matemática y el valor verdadero del conjunto de atributos que indica su nivel informativo con respecto a la clase. 

Dada una muestra de $m$ casos tomada de una población $U$, llamamos estadístico a todo valor $M$ computado a partir de los datos presentes en la muestra. Es decir, un estadístico es una función que mapea el espacio $\mathcal{E}$ de posibles muestras de tamaño $m$ al conjunto de los números reales ($M: \mathcal{E} \to \mathbb{R}$). Si la muestra es tomada en forma aleatoria, entonces $M$ es una variable aleatoria y por tanto posee una función de probabilidad $P(M)$ y un valor esperado $E(M)$. Consecuentemente, medidas de evaluación como $SU$ y $MSU$, calculadas sobre una muestra aleatoria, son a su vez variables aleatorias y cada una, considerada sobre un grupo de atributos, posee una esperanza matemática. 

\begin{definition} \label{def_sesgo}
Dado el conjunto $\left\lbrace A_1, A_2,\dots,A_n, C\right\rbrace$ conformado por n atributos discretos o categóricos junto con la clase $C$, un tamaño muestral m, una medida de evaluación ${M(a_{m1},a_{m2},\dots ,a_{mn},c_{m})\,}$ y una constante $\theta$ que indica el valor verdadero del nivel informativo en conjunción de los atributos acerca de la clase, el sesgo de evaluación de atributos está dado por
	\begin{equation}
        Sesgo_\theta \left[M\right] = E_\theta(M)-\theta.
    \end{equation}
donde $E_\theta(M)$ denota el valor esperado de la medida de evaluación $M$ dado el valor verdadero $\theta$ que se estima.
\end{definition}

Además de las medidas multivariadas, la Definición \ref{def_sesgo} es aplicable a medidas bivariadas como el $SU$ donde el conjunto estará conformado por un único atributo junto con la clase. Finalmente para el caso en que $M$ sea univariada, la Definición \ref{def_sesgo} se reduce a la definición clásica de sesgo (\textit{bias}) que se ofrece en probabilidad y estadística.

%Además de las medidas multivariadas, la Definición \ref{def_sesgo} es aplicable a medidas bivariadas como el $SU$ donde el conjunto estará conformado por un único atributo junto con la clase. 
%En estadística se llama sesgo de un estimador a la diferencia entre su esperanza matemática y el valor numérico del parámetro que estima. Un estimador cuyo sesgo es nulo se llama insesgado o centrado.
%\begin{definition}
%En notación matemática, dada una muestra ${\displaystyle X_{1},\dots ,X_{n}\,iidX}$ y un estimador ${\displaystyle T(x_{1},\dots ,x_{n})\,}$ del parámetro poblacional $\theta\,$, el sesgo es:
%${\displaystyle E(T)-\theta \,} {\displaystyle E(T)-\theta \,}$
%\end{definition}

%Se llama parámetros poblacionales a cantidades que se obtienen a partir de las observaciones de la variable y sus probabilidades y que determinan perfectamente la distribución de esta, así como las características de la población, por ejemplo: La media, μ, la varianza σ2, la proporción de determinados sucesos, P.
%Los Parámetros poblacionales son números reales, constantes y únicos.

\section{Resumen}
En este capítulo se ha expuesto los fundamentos teóricos tanto de la teoría de la información como así también del proceso de selección de atributos. 

Cabe destacar, que las Definiciones propias introducidas de Cardinalidad Univariada \ref{card_uni}, Cardinalidad Multivariada \ref{card_multi} Sesgo en la Evaluación de Atributos \ref{def_sesgo} establecen el marco formal correspondiente para el análisis de sesgo llevado a cabo en la presente tesis.

  \SetKwInput{Kw}{Entrada} 
\chapter{ANALISIS DE SESGO Y PROPUESTA}

En esta sección explicaremos el análisis de sesgo realizado sobre el $MSU$, como así también los factores identificados y que tendrían implicancia directa en el comportamiento de dicha medida basada en la teoría de la información.

Como resultado del mencionado análisis, proponemos una relación de asociación empírica entre los factores identificados que permiten un comportamiento controlado del $MSU$.

Finalmente, se analiza la relación de asociación propuesta desde el punto de vista de la inferencia estadística mediante la simulación de pruebas de bondad de ajuste \cite{Riedwyl}.

\section{Propósito de los experimentos}
El escenario experimental cuyos detalles se explican en el siguiente apartado, ha sido montado con el objetivo de analizar el comportamiento del $SU$ y el $MSU$ basado específicamente en la cardinalidad de las variables (atributos y de la clase), el tamaño muestral y la cantidad de variables que conforma el conjunto a ser evaluado.

\section{Escenario Experimental}
En el presente trabajo, hemos adoptado la técnica de simulación Monte Carlo del escenario propuesto en \cite{liu} al cual hemos incluido, la creación por un lado de atributos individualmente informativos por medio del método de Kononenko \cite{kononenko} y por otra parte la inclusión de atributos colectivamente informativos por medio de la disyunción exclusiva (XOR).

Los resultados de cada experimento ha sido promediado sobre un total de $1000$ corridas. 

\subsection{Datos sintéticos}
Los conjuntos de datos generados, presentan las siguientes características:
\begin{enumerate}
\item Una variable de clasificación (``la clase'') con $2$ ó $10$ valores posibles.
\item Atributos informativos y non-informativos con cardinalidades univariadas de $2$, $4$, $5$, $8$, $10$, $16$, $20$, $30$, $32$, $40$ y $64$.
\item Los atributos no-informativos fueron creados aleatoriamente a partir de la distribución uniforme de forma independiente a la clase.
\item Los atributos individual e igualmente informativos fueron creados utilizando el Método de Kononenko $(MK)$ el cual se describe en el siguiente apartado.
\item Los atributos que en su conjunto son informativos, fueron creados mediante la función Ó-Exclusivo $(XOR)$, al cual para hacerlo más real se ha agregado ruído de la forma $P(clase = XOR(f_1,f_2)) = 0.95,P(class \not= XOR(f_1,f_2)) = 0.05$. 
\item  En un conjunto de experimentos, se ha utilizado tamaños muestrales fijos de $1000$ y $5000$ instancias y en otro conjunto de experimentos el tamaño muestral es variable.
\end{enumerate}

\subsubsection{Método de Kononenko (MK)}

Este método permite crear atributos de diferentes cardinalidades pero que sin embargo, contienen una misma cantidad de información acerca de la $clase$. 

Esto es posible gracias a la unión de los posibles valores de un atributo en dos subconjuntos, el primero con $\left\lbrace1, ..., (V\;div\;2)\right\rbrace$ y el segundo con $\left\lbrace (V\:div\;2\;+\;1), ..., V \right\rbrace$ para un total de $V$ valores posibles. 

La probabilidad que el valor provenga de uno de los dos subconjuntos está dada por la clase, mientras que la selección de un valor dentro de un subconjunto particular es aleatoria a partir de la distribución uniforme.

La probabilidad que el valor del atributo se encuentre en uno de los dos subconjuntos, está definido por:
\begin{equation}
  P\bigg(j \in \left\lbrace1,...,(\lfloor\frac{V}
  {2}\rfloor)\right\rbrace|i\bigg) :=
      \left\{ \begin{array}{ll}
              1/(i+kC)    & \mbox{si $i$ mod 2 = 0} \\
              1 -1/(i+kC) & \mbox{si $i$ mod 2 $\ne$ 0}
               \end{array} \right.
\end{equation}
donde $C$ es la cardinalidad de la clase, $i$ es un entero que indexa el valor posible de la clase $\left\lbrace c_1,...,c_i\right\rbrace$ y $k$ determina cuan informativo será el atributo. 

Un alto valor de $k$ indica una fuerte relación de asociación entre el atributo y la clase lo cual lo convierte en un atributo más informativo. En el presente trabajo, para todos los experimentos se ha utilizado $k = 1$.

\subsection{Entorno Computacional}
A continuación se describe brevemente los detalles del entorno de ejecución de los experimentos realizados:

\begin{itemize}
\item Los algoritmos fueron implementados mediante el lenguaje de programación R en su versión  3.3.2 de ``The R Foundation for Statistical Computing".
\item Distribución ``Arch Linux" con plataforma GNU/Linux x86\_64 y versión de kernel 4.8.13-1.
\item Procesador Intel(R) Core(TM) i7\-4500U CPU @ 1.80GHz.
\item Memoria RAM de 8 GB.
\end{itemize}

\section{IG, SU y MSU}
Las medidas basadas en información como el $IG$ tienden a sobrestimar o favorecer a los atributos con mayor cantidad de valores posibles, es decir que dichos atributos aparecerán sistemáticamente como más predictivos o más correlacionados con la clase. Esto es lo que se conoce como sesgo en la medida de evaluación de atributos \ref{def_sesgo}.

Trabajos anteriores han demostrado que ante un incremento de la cardinalidad univariada \ref{card_uni} se produce un suave decremento exponencial del $SU$ para atributos informativos y un crecimiento lineal para atributos no-informativos \cite{hall}, esta situación lo hemos verificado nuevamente con el fin de analizar el comportamiento del $MSU$ como una generalización del $SU$. 

\subsection{Atributos informativos y no-informativos}
\subsubsection{Efecto de la cardinalidad univariada}
\begin{figure}[!htb]
    \subfloat[Atributos no-informativo e individualmente informativo. La cardinalidad de la clase es $10$. 
    \label{subfig:a1}]{
      \includegraphics[width=0.5\textwidth]{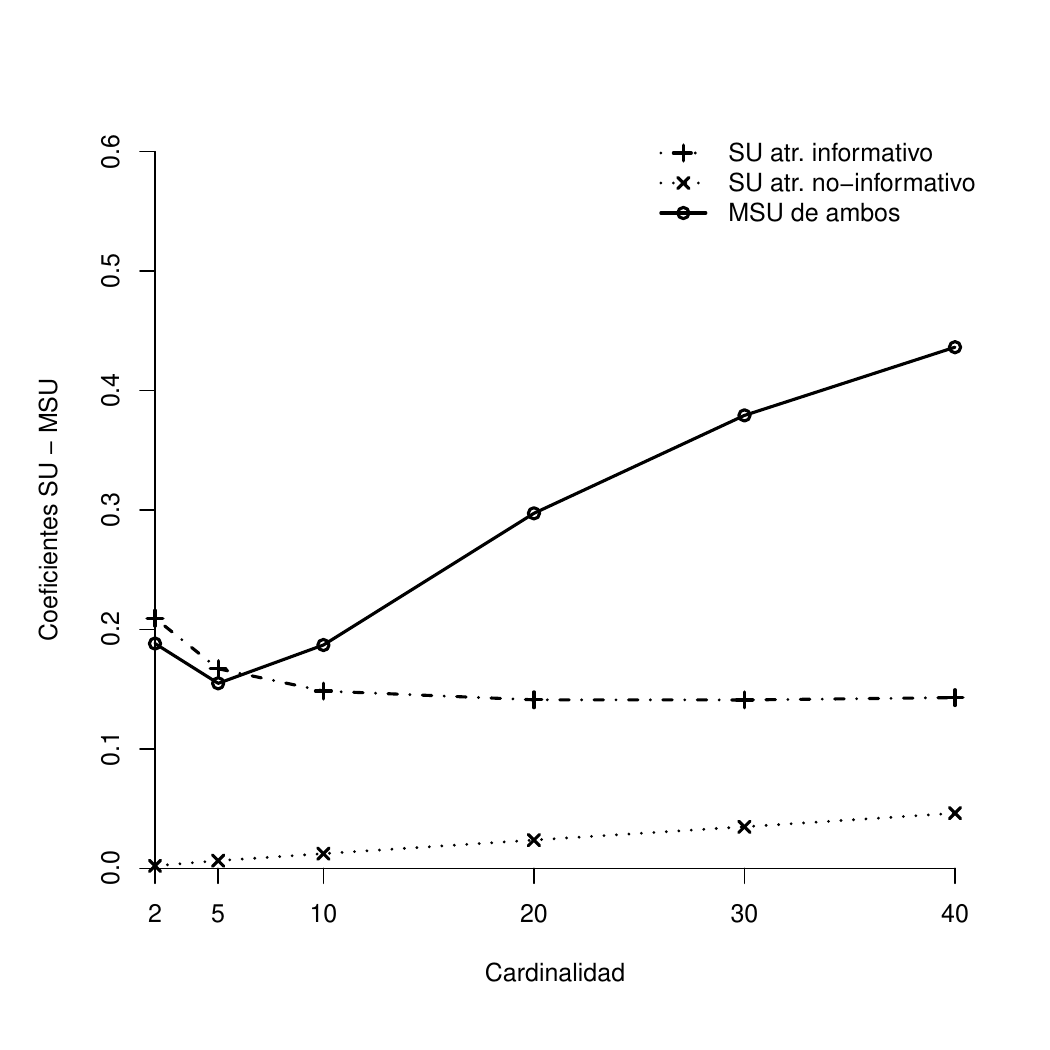}
    }
    \quad
    \subfloat[Atributos no-informativo e individualmente informativo. La cardinalidad de la clase es $2$. 
\label{subfig:a2}]{
      \includegraphics[width=0.5\textwidth]{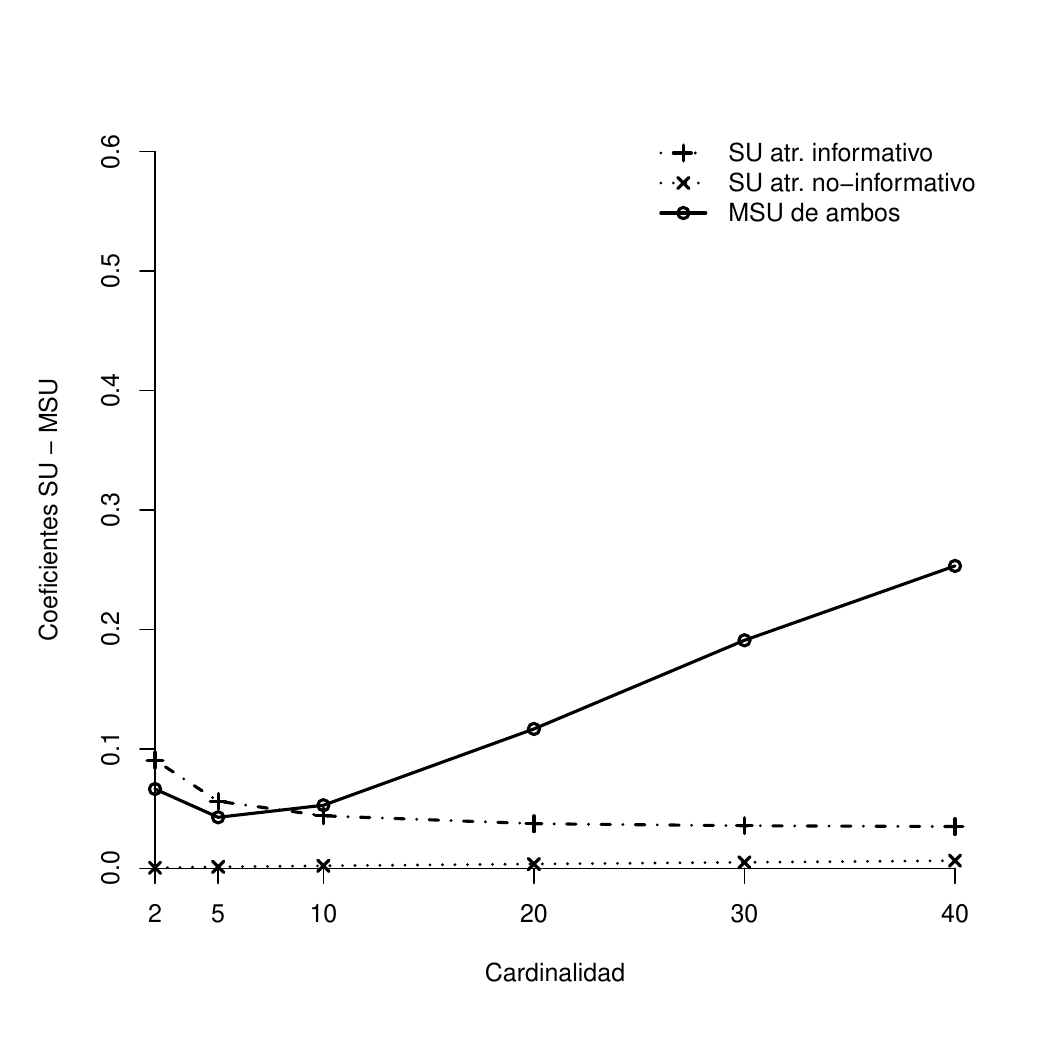}
    }
    \caption{Los efectos de la variación de la cardinalidad sobre el SU y el MSU para un tamaño muestral de $1000$ instancias.}
    \label{fig:figuras-a1-a2}
\end{figure}

Para el $MSU$ la interacción de éstos tipos de atributos muestra un decrecimiento inicial (siguiendo la tendencia del atributo informativo) seguido de un crecimiento estable (como el atributo no-informativo) tal como se puede observar en las Figuras \ref{fig:figuras-a1-a2}(a) y (b).

Por otra parte, se puede apreciar que si bien el número de clases hace que los valores obtenidos por el $SU$ y el $MSU$ sean ligeramente menores, el comportamiento es el mismo ante el incremento de la cardinalidad de los atributos para un tamaño muestral fijo. 

Por lo tanto, se puede concluir en primera instancia, que el sesgo del $MSU$ para una combinación de atributos informativos y no-informativos se encuentra asociado al sesgo individual de cada atributo del conjunto y este a su vez en base a la cardinalidad univariada.

\subsubsection{Efecto del tamaño muestral}

\begin{figure}[!htb]
    \subfloat[Subconjuntos de atributos individualmente informativos y no-informativos con tamaño muestral de $8$ a $50$ instancias. \label{subfig:e2}]{
      \includegraphics[width=0.5\textwidth]{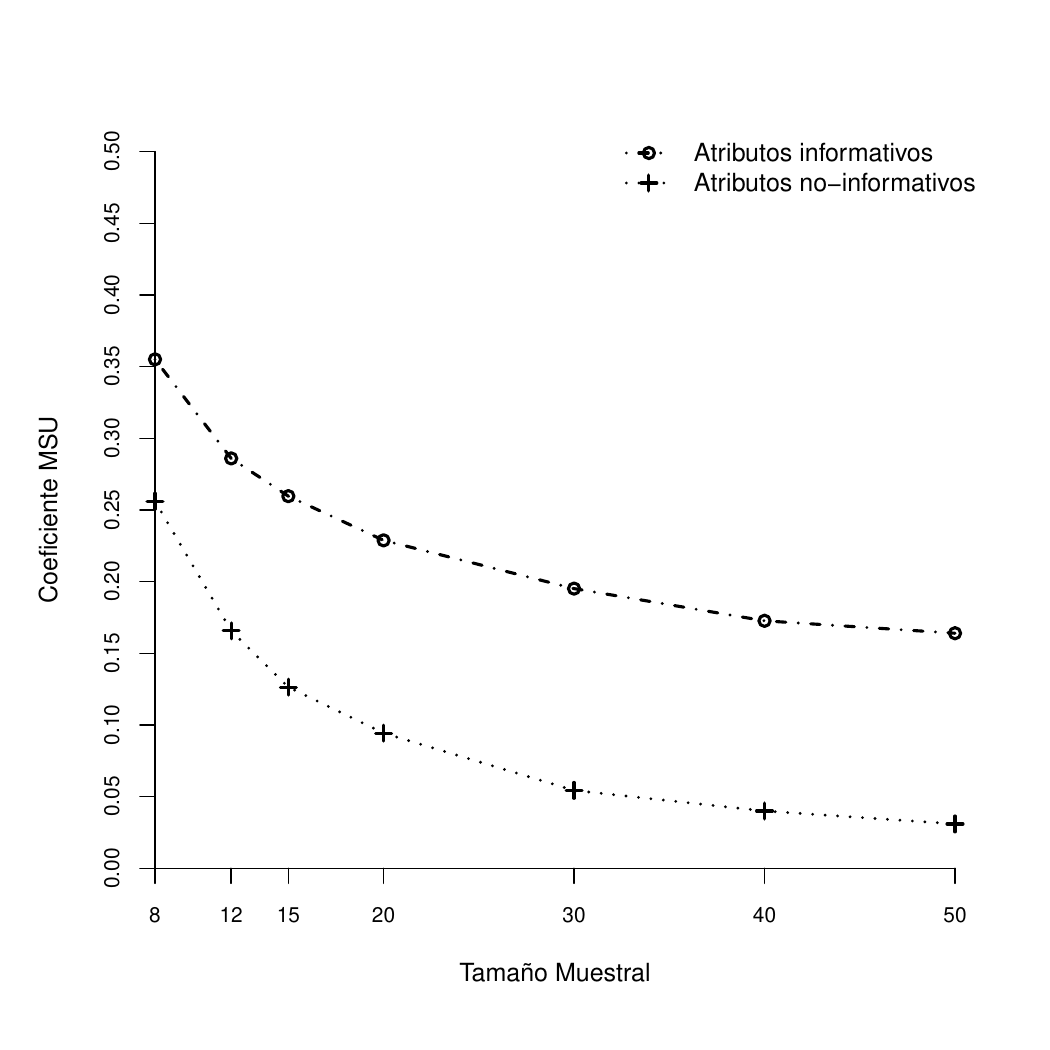}
    }
    \quad
    \subfloat[Subconjuntos de atributos individualmente informativos y no-informativos con tamaño muestral de $8$ a $150$ instancias. \label{subfig:e}]{
      \includegraphics[width=0.5\textwidth]{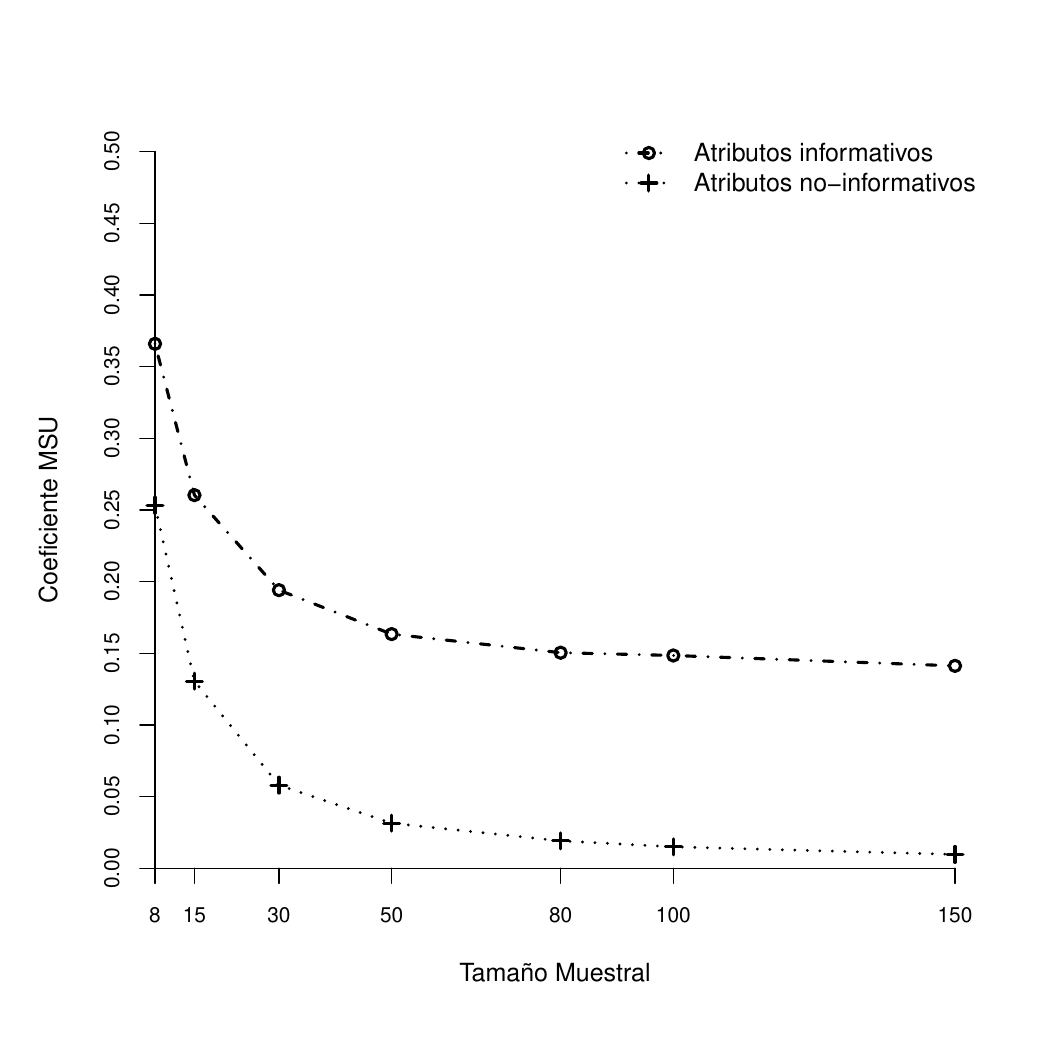}
    }
    \caption{Los efectos de la variación del tamaño muestral sobre el MSU para subconjuntos de atributos individualmente informativos y no-informativos. La cardinalidad de los atributos y la clase es de $2$. El tamaño de los subconjuntos es de $3$ elementos incluida la clase.}
    \label{fig:figuras-e1-e2}
\end{figure}

Para un subconjunto $A$ conformado por $2$ atributos individualmente informativos y la clase (todos con cardinalidad igual a $2$), en las Figuras \ref{fig:figuras-e1-e2}(a) y (b), se puede apreciar un decrecimiento suave en el coeficiente del $MSU$ como resultado del efecto de la variación del tamaño muestral. Cabe destacar, que este decremento tiende a estabilizarse en un valor mayor a $cero$ como se puede apreciar en la Figura \ref{fig:figuras-e1-e2}(b), lo cual es lo esperado por la naturaleza informativa de los atributos.  

Por otra parte, dado un subconjunto $A$ conformado por $2$ atributos no-informativos y la clase (todos con cardinalidad igual a $2$), en las Figuras \ref{fig:figuras-e1-e2}(a) y (b), se puede apreciar un decrecimiento suave en el coeficiente del $MSU$ como resultado del efecto de la variación del tamaño muestral. Cabe destacar, que este decremento tiende a estabilizarse en un valor igual a $cero$ como se puede apreciar en la Figura \ref{fig:figuras-e1-e2}(b), lo cual es lo esperado por la naturaleza no-informativa de los atributos.  

Por lo tanto, se puede concluir parcialmente, que a partir de cierto tamaño muestral para cardinalidades fijas, el $MSU$ se estabiliza en un valor coherente con la naturaleza de los atributos que conforman el conjunto evaluado. 

\subsection{Atributos colectivamente informativos}

El $SU$ es una medida \textit{bivariada}, es decir, solo puede medir la correlación entre un atributo y una clase o entre dos atributos entre sí.

Esta limitación impide que el $SU$ pueda tomar en cuenta los casos donde
se necesita más de una variable (atributo) para determinar un concepto (clase), como en el caso de $XOR$.

\begin{figure}[!htb]
    \subfloat[Atributos colectivamente informativos con tamaño muestral de $8$ a $50$ instancias. \label{subfig:b2}]{
      \includegraphics[width=0.5\textwidth]{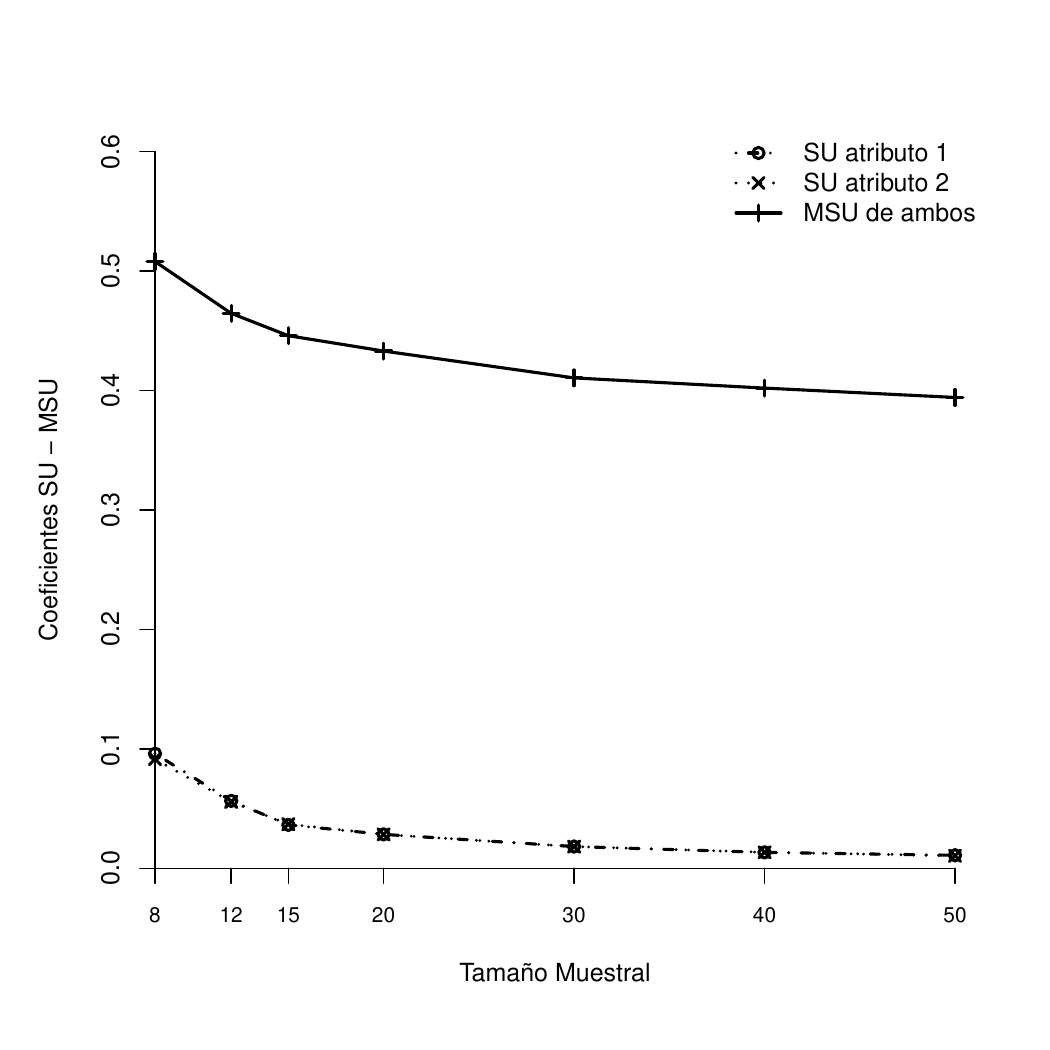}
    }
    \quad
    \subfloat[Atributos colectivamente informativos con tamaño muestral de $8$ a $150$ instancias. \label{subfig:b}]{
      \includegraphics[width=0.5\textwidth]{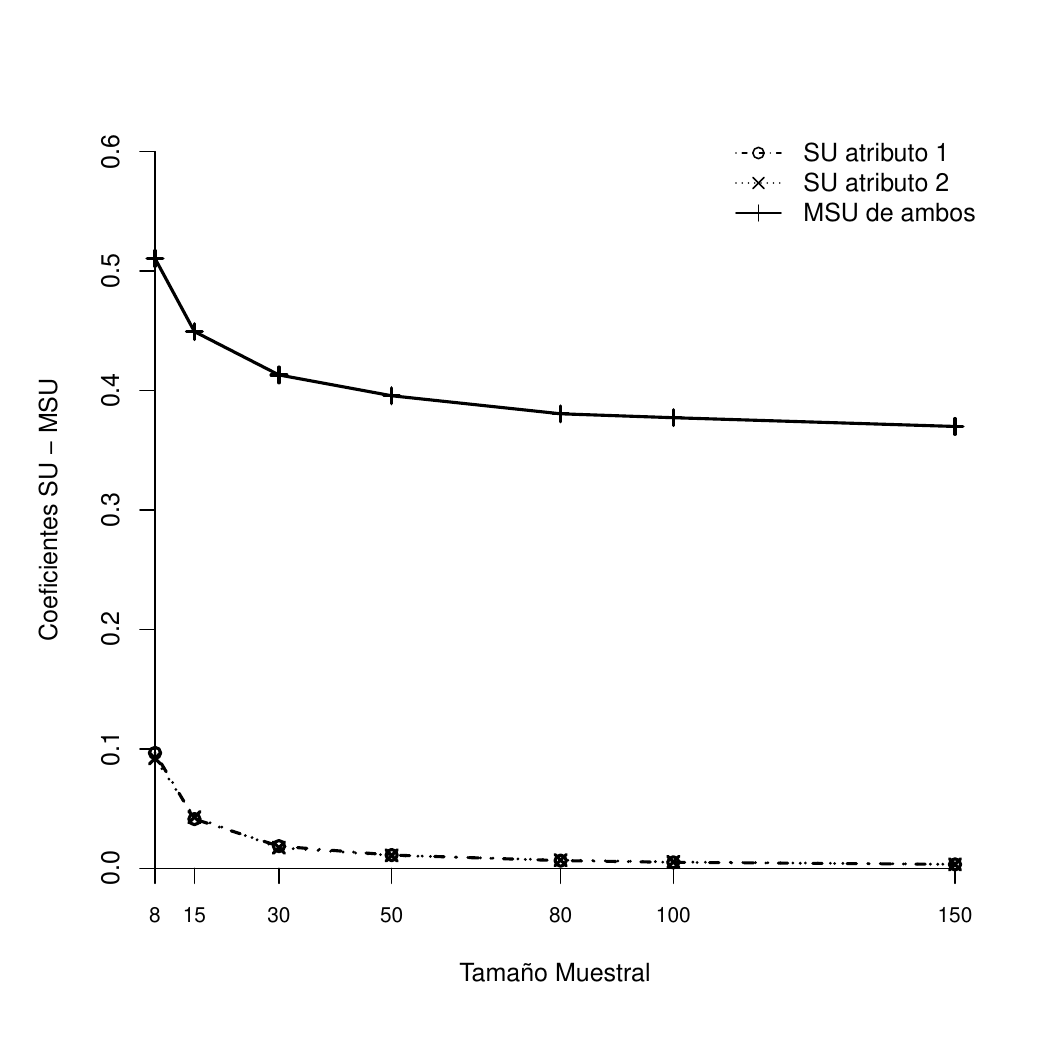}
    }
    \caption{Los efectos de la variación del tamaño muestral sobre el SU y el MSU para atributos colectivamente informativos mediante XOR con un ruido del $5\%$. La cardinalidad de los atributos y la clase es de $2$. }
    \label{fig:figuras-b1-b2}
\end{figure}

Esta limitación es resuelta por el $MSU$ tal como puede observarse en las Figuras \ref{fig:figuras-b1-b2}(a) y (b) donde el $SU$ considera a los atributos individuales como no-informativos mientras que el $MSU$ establece que los mismos en conjunto efectivamente son informativos. 

Cabe destacar que este tipo de juzgamiento de atributos es de una gran importancia en el proceso de selección de atributos.

Por otra parte, en las Figuras \ref{fig:figuras-b1-b2}(a) y (b), se puede apreciar además un suave decrecimiento en los coeficientes como resultado del efecto del tamaño muestral tanto sobre el $SU$ como así también sobre el $MSU$.

Por lo tanto, se puede concluir parcialmente, que a partir de cierto tamaño muestral para cardinalidades fijas, las medidas se estabilizan manteniendo su valor. 

\section{Sesgo en el MSU}

En \cite{quinlan_r_r} se muestra que la ganancia de un atributo $A$ (medida con respecto a la clase u otro atributo) es menor o igual a la ganancia de un atributo $A'$ formado dividiendo aleatoriamente $A$ en un mayor número de valores posibles (cardinalidad univariada). 

Esto significa que, en general, el atributo derivado (y por analogía, atributos con más valores posibles) parecerá ser más predictivo o correlacionado con la clase que el original.

Al error sistemático que se produce de igual modo en todas las mediciones de la correlación entre las variables (atributos entre sí o atributos y clase) es lo que se conoce como Sesgo \ref{def_sesgo}.

Puesto que el $MSU$ es una generalización del $SU$ y este a su vez se basa en la normalización del $IG$, para llevar a cabo un análisis del sesgo del $MSU$, hemos extendido la referida demostración\cite{quinlan_r_r} el cual puede ser observado en la Tabla \ref{tabla:MSU}, para el cual se ha generado los valores tanto de los \textit{atributos} como así también de la \textit{clase} de forma aleatoria independientemente entre sí (atributos no-informativos) a partir de la distribución uniforme.

\begin{table}[!htb]
  \centering 
  \subfloat[][$MSU=0$]{
    \label{tab:a}
      \begin{tabular}{c|c||c}
          $f_1$ & $f_2$ & $clase$\\
          \hline
          $b$ & $s$ & $p$ \\
          $b$ & $s$ & $q$ \\
          $b$ & $t$ & $p$ \\
          $b$ & $t$ & $q$ \\
          $a$ & $s$ & $p$ \\
          $a$ & $s$ & $q$ \\
          $a$ & $t$ & $p$ \\
          $a$ & $t$ & $q$ \\
      \end{tabular}
    }
    \qquad
  \subfloat[][$MSU=0.10$]{
    \label{tab:b}
      \begin{tabular}{c|c||c}
          $f_1'$ & $f_2$ & $clase$\\
          \hline
          \boldmath$a$ & $s$ & $p$ \\
          $b$ & $s$ & $q$ \\
          $b$ & $t$ & $p$ \\
          $b$ & $t$ & $q$ \\
          $a$ & $s$ & $p$ \\
          $a$ & $s$ & $q$ \\
          $a$ & $t$ & $p$ \\
          $a$ & $t$ & $q$ \\
      \end{tabular}
    }
    \qquad
  \subfloat[][$MSU=0.18$]{
    \label{tab:c}
      \begin{tabular}{c|c||c}
          $f_1''$ & $f_2$ & $clase$\\
          \hline
          \boldmath$c$ & $s$ & $p$ \\
          $b$ & $s$ & $q$ \\
          $b$ & $t$ & $p$ \\
          $b$ & $t$ & $q$ \\
          $a$ & $s$ & $p$ \\
          $a$ & $s$ & $q$ \\
          $a$ & $t$ & $p$ \\
          $a$ & $t$ & $q$ \\
      \end{tabular}
    }
    \caption{Efectos de las cardinalidades sobre el MSU para atributos no-informativos.}
    \label{tabla:MSU}
\end{table}

En la tabla \ref{tab:a} se puede observar que tanto los atributos $f_1$ y $f_2$, como así también la $clase$ poseen cardinalidad $2$, es decir, que puede adquirir $2$ valores posibles. En este caso, para un total de $8$ instancias con los valores distribuidos equitativamente se puede apreciar que el $MSU$ del conjunto es $cero$, es decir, las $3$ variables no se encuentran correlacionadas entre sí, lo cual es lo esperado.

Luego, manteniendo la cardinalidad de las variables, hemos seleccionado aleatoriamente una variable y una instancia de dicha variable para invertir su valor como puede apreciarse en la tabla \ref{tab:b}. Esto ha ocasionado que el $MSU$ incremente su valor indicando sesgada y erróneamente que existe una correlación entre las variables.

A partir de este hecho, podemos observar que al invertir el valor de una variable, hemos creado una combinación (instancia) redundante de valores por un lado y por el otro lado, ha quedado una combinación (instancia) única de valores generando una correlación por unicidad.

Este hallazgo, nos permite establecer un factor que influye en el sesgo del $MSU$ al cual hemos denominado \textit{cardinalidad multivariada} y lo hemos definido formalmente en \ref{card_multi}.

Cabe destacar, que en caso de eliminar al menos una instancia del conjunto de datos, estaríamos afectando de la misma forma al $MSU$ ya que el \textit{tamaño muestral} no sería suficiente para albergar la \textit{cardinalidad multivariada}, es decir, la cantidad de posibles combinaciones de los valores de cada atributo que conforma el conjunto incluida la clase.  

Sin embargo, si en lugar de invertir el valor como en el caso anterior, aumentamos la cardinalidad univariada de $f_1$ agregando una nuevo valor posible a su alfabeto tal como se puede apreciar en la Tabla \ref{tab:c}, entonces se puede notar el $MSU$ se ve afectado nuevamente denotando una correlación entre las variables del conjunto, esto es, exactamente como ocurre con el $SU$ y que fuera estudiado en \cite{hall}. 

%En esta tesis, al mencionado alfabeto le hemos denominado \textit{cardinalidad univariada} cuya definición formal lo hemos expresado de la siguiente forma:
%
%\begin{definition*} \label{def:card:univariada}
%Dada una característica discreta o categórica $A$, su Cardinalidad Univariada, denotada por $\left\vert{A}\right\vert$, es el número posible de etiquetas diferentes que puede tomar $A$.
%\end{definition*}

Finalmente, puede observarse fácilmente que el efecto de ampliar el conjunto de valores posibles de una variable impacta de forma más significativa al $MSU$ (con respecto a lo expuesto en \ref{tab:b}) puesto que estamos alterando ambas cardinalidades.

\subsection{Relación entre cardinalidades}
\begin{figure}[!htb]
    \subfloat[Mapeo entre la cardinalidad univariada y la multivariada para atributos informativos.
      \label{subfig:c}]{%
      \includegraphics[width=0.5\textwidth]{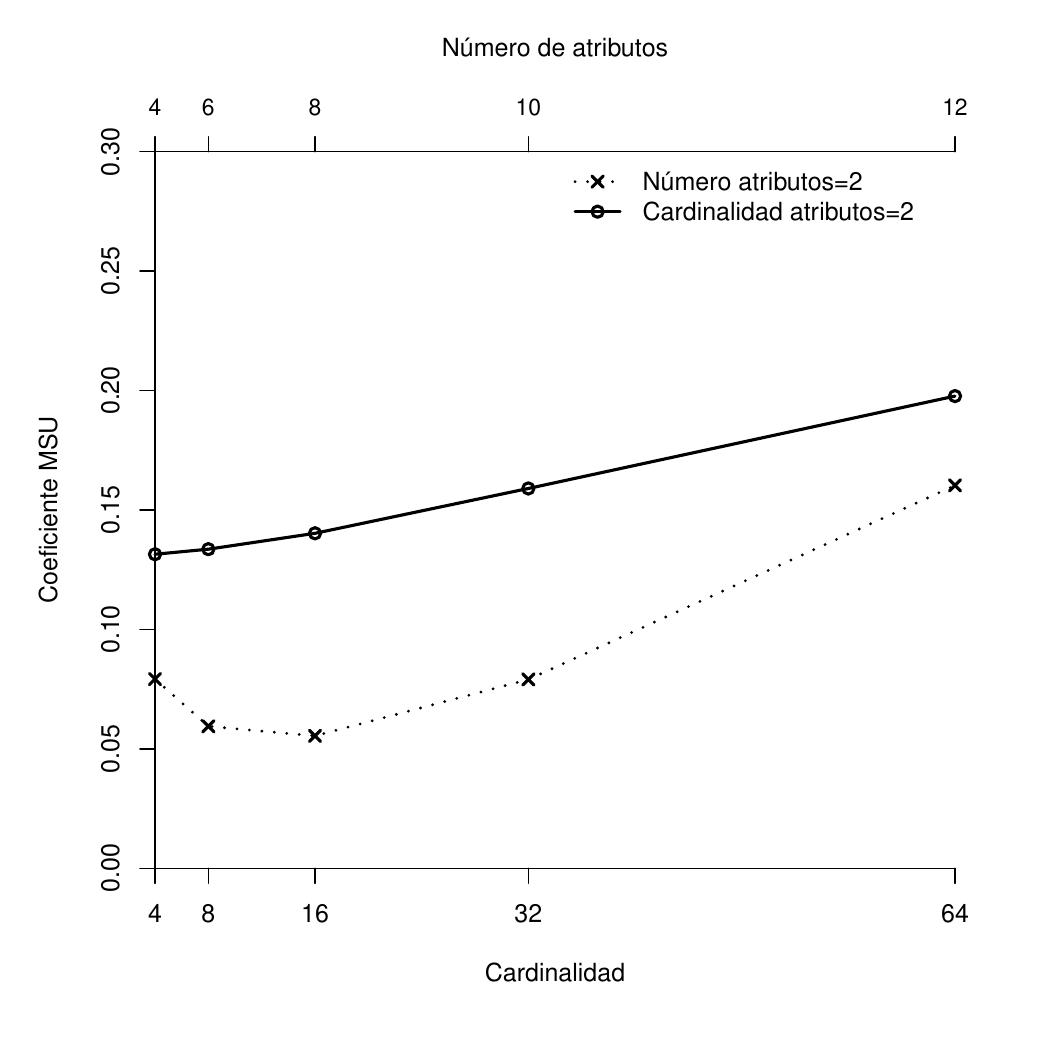}
    }
    \quad
    \subfloat[Mapeo entre la cardinalidad univariada y la multivariada para atributos informativos. 
      \label{subfig:d}]{%
      \includegraphics[width=0.5\textwidth]{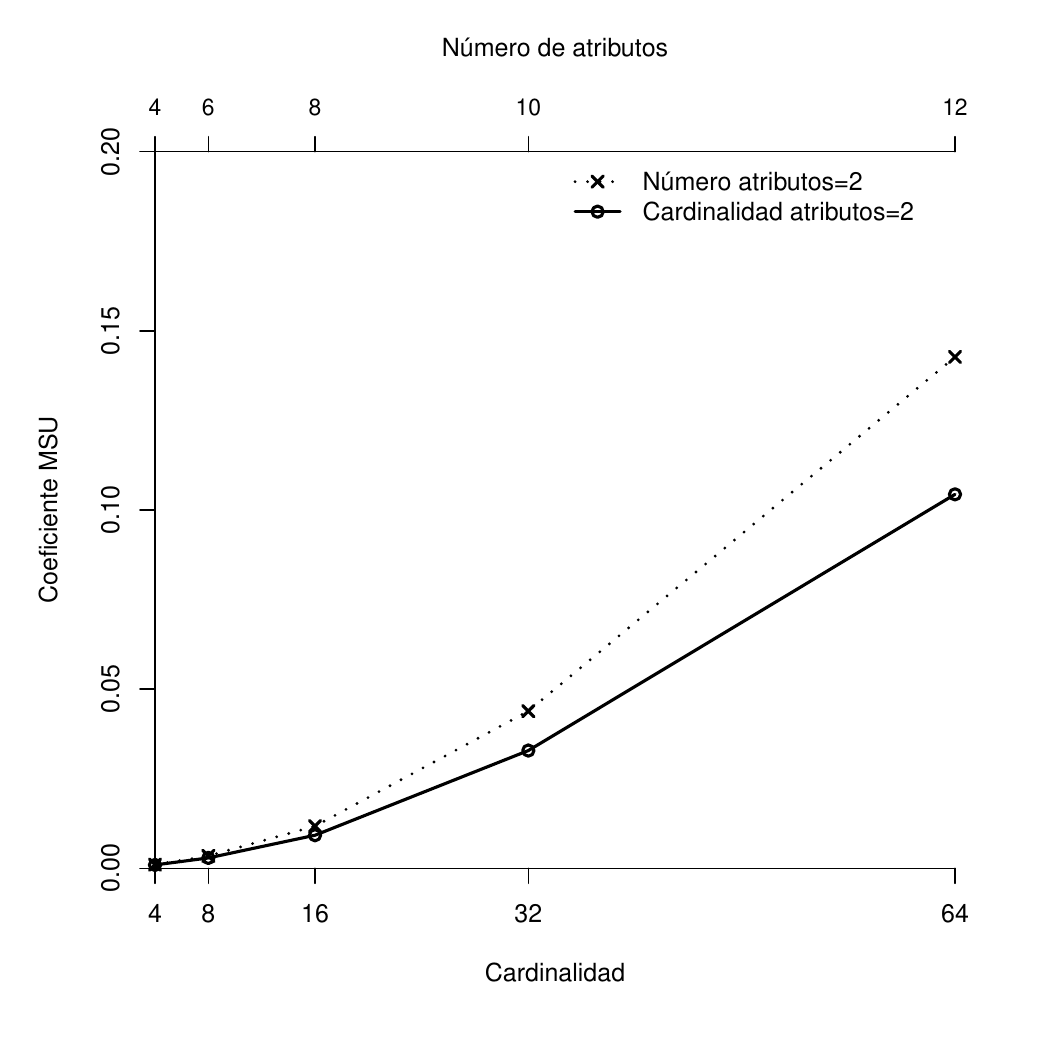}
    }
    \caption{Los efectos de la variación de las cardinalidades univariadas y multivariadas sobre el MSU. La cardinalidad de la clase es $2$ y el tamaño muestral es de $5000$ instancias.}
    \label{fig:figuras-cd}
\end{figure}

De tal forma a sustentar nuestra definición propuesta de la cardinalidad multivariada \ref{card_multi} como factor de sesgo en el $MSU$ y atendiendo el conocido sesgo provocado por la cardinalidad univariada \ref{card_uni} en las medidas basadas en información, hemos llevado a cabo un estudio como puede apreciarse en la Figura \ref{fig:figuras-cd}, donde demostramos una relación de equivalencia entre ambas cardinalidades tanto para atributos informativos, como así también para atributos no-informativos que conforman el conjunto evaluado por el $MSU$.

Por ejemplo, dados $2$ subconjuntos $A$ y $B$ cada una con una $clase$ con cardinalidad univariada igual a $2$, para establecer una equivalencia de cardinalidades entre los atributos de los subconjuntos, si el subconjunto $A$ está conformado por $2$ atributos $a_1$ y $a_2$ cada uno con cardinalidad univariada igual a $4$, entonces el subconjunto $B$ debe estar conformado por $4$ atributos $b_1$, $b_2$,$b_3$ y $b_4$ cada uno con cardinalidad univariada igual a $2$. 

De esta forma, la cardinalidad multivariada de los subconjuntos $A$ y $B$, para el ejemplo mencionado viene dado por 
\begin{equation*}
\begin{split}
\left\vert{clase}\right\vert\prod_{i=1}^{2} \left\vert{a_{i}}\right\vert & = 
\left\vert{clase}\right\vert\prod_{i=1}^{4} \left\vert{b_{i}}\right\vert \\
2 * 4 * 4 & = 2* 2 * 2 * 2 * 2. \\ 
\end{split}
\end{equation*}

\subsection{Factores de Sesgo en el MSU}
Basado en el análisis del comportamiento del $MSU$ realizado y descrito en la sección anterior podemos establecer que los factores implicados en el error sistemático o sesgo en la medida $MSU$ son:

\begin{description}
\item[Cardinalidad Univariada] debido a que el $MSU$ es una medida basada en la teoría de la información.
\item[Cardinalidad Multivariada] debido a que el $MSU$ es una medida capaz de evaluar la interacción en conjunto de las variables.
\item[Tamaño Muestral] debido a que el $MSU$ es un estadístico muestral con la propiedad de consistencia, es decir, el mismo converge a su valor verdadero cuando el número de datos de la muestra tiende a infinito o al tamaño total de la población.
\end{description}

\subsection{Relación de Asociación Propuesta}

Dado el análisis de los factores de sesgo en el $MSU$, podemos deducir que la cardinalidad multivariada es el factor que engloba a los demás factores ya que el mismo se calcula en base a la cardinalidad univariada y por otra parte, como habíamos demostrado anteriormente, el tamaño de la muestra debe ser tal que ella albergue al menos todas las combinaciones posibles de valores de las variables del conjunto.

Por lo tanto, proponemos que el tamaño muestral esté dado en función a un múltiplo de varias veces la cardinalidad multivariada, de tal forma a facilitar una alta probabilidad de \textit{total representatividad} en la muestra.

%el tamaño muestral debe poder contemplar al menos todas las combinaciones posibles de las variables del conjunto.
%En efecto, y haciendo una analogía, la cardinalidad multivariada es el número de individuos diferentes que integran una población y por lo tanto, proponemos que el tamaño muestral debe estar dado en función a un múltiplo de veces la cardinalidad multivariada, esto es, de tal forma a garantizar que exista la probabilidad de \textit{total representatividad} en la muestra.  

En este sentido, los experimentos realizados reflejan una clara tendencia para controlar el sesgo en el $MSU$ dada la siguiente relación de asociación que proponemos 
\begin{equation}
Tama\tilde{n}o\ muestral \approx 10\left\vert{clase}\right
\vert\prod_{i=1}^{n} \left\vert{f_{i}}\right\vert.
\end{equation}

Esta aproximación empírica, se basa en los hallazgos descritos a continuación:
\begin{itemize}
\item Para atributos \textit{colectivamente informativos}, en la Figura \ref{fig:figuras-b1-b2} puede observarse que el $MSU$ se estabiliza (con variabilidad menor al $1\%$) a partir del tamaño muestral $80$ lo que representa $10$ veces la cardinalidad multivariada.
\item Para atributos \textit{individualmente informativos}, en la Figura \ref{subfig:c} para $5000$ instancias (aproximadamente $10$ veces la cardinalidad multivariada), se puede observar un punto de inflexión para el subconjunto de $8$ atributos con cardinalidad univariada igual a $2$ y el cambio del ritmo de crecimiento para el subconjunto de $2$ atributos con cardinalidad univariada igual a $16$. 

Ademas, otra evidencia de este comportamiento se puede apreciar en la Figura \ref{fig:figuras-e1-e2} (b) donde el $MSU$ se estabiliza (con variabilidad menor al $1\%$) a partir del tamaño muestral $80$ el cual coincide con la aproximación propuesta. 
\item Para atributos \textit{no-informativos}, en la Figura \ref{subfig:d} para $5000$ instancias (aproximadamente $10$ veces la cardinalidad multivariada), se puede observar un punto de separación para ambos subconjuntos a partir de $8$ atributos con cardinalidad univariada igual a $2$ para un subconjunto y de $2$ atributos con cardinalidad univariada igual a $16$ para el otro subconjunto. 

Además, otra evidencia de este comportamiento se puede apreciar en la Figura \ref{fig:figuras-e1-e2} (b) donde el $MSU$ se estabiliza (con variabilidad menor al $1\%$) a partir del tamaño muestral $80$ el cual coincide con la aproximación propuesta.
\end{itemize}

\subsection{Tamaño muestral para la total representatividad}

En base al análisis descrito en la sección anterior, en donde se busca controlar el sesgo en el $MSU$ por medio de un tamaño muestral que garantice una total representatividad en la muestra, en esta sección llevaremos a cabo la simulación de una prueba de bondad de ajuste \cite{Riedwyl} con el fin de obtener una aproximación de referencia desde la perspectiva de la inferencia estadística.

En este sentido, considérese el conjunto de $3$ variables aleatorias con distribución de Bernoulli, las cuales representan a los atributos incluyendo a la clase, todos con cardinalidad univariada igual a $2$. Por lo tanto, el espacio muestral estará compuesto por $8$ combinaciones posibles (esto es, su cardinalidad multivariada). 

Luego, cada resultado obtenido estará compuesto por $3$ valores (uno por parte de cada variable) donde dichos resultados están distribuidos como una densidad multinomial con probabilidades $p_1$, ..., $p_8$ supuestas por la probabilidad individual o frecuencias de los valores presentes en cada combinación. 

Por simplicidad, asumimos que cada resultado sea equiprobable con $1/8$ de probabilidad. Asumimos también que cada variable será representada por una moneda corriente, de tal forma a que cada resultado viene a ser físicamente simulable por el lanzamiento simultáneo de $3$ monedas que sean distinguibles. 

Ahora supongamos que obtenemos una muestra de $m$ resultados de lanzamientos. Mediante el test de bondad de ajuste $\chi^2$ (chi-cuadrado) \cite{cressie} \cite{stark} es posible verificar la hipótesis nula de que los resultados de la muestra obtenida mediante este procedimiento son simplemente una consecuencia de las probabilidades establecidas. 

Sin embargo, para que se cumpla el objetivo de una muestra totalmente representativa, es requerido que cada resultado posible ocurra al menos una vez en los $m$ lanzamientos, esto es, no debe existir ningún resultado posible con frecuencia observada igual a cero. 

Para representar esta simulación, en la Tabla \ref{tab:chisq} se muestran las columnas de \textit{Resultado Posible}, \textit{Frecuencia Observada}, \textit{Frecuencia Esperada} y el \textit{Término ${\chi}^2$}, los cuales corresponden al espacio muestral, la frecuencia directamente observada, la frecuencia teórica esperada y el cálculo del estadístico $chi$-$cuadrado$ respectivamente.

En este sentido, consideramos que la tarea de simulación consiste en la escritura de valores regulares de muestra, incluyendo un cero en la columna de \textit{Frecuencia Observada} de tal forma a comparar el valor calculado resultante del $\chi^2$ y el valor crítico del mismo, el cual depende únicamente del nivel de significación $\alpha$ y de los grados de libertad (los cuales permanecen fijos con valores de $0.05$ y $7$ respectivamente). 

De esta forma podemos evaluar la probabilidad de cualquier conjunto de $m$ resultados donde a medida que incrementamos $m$, viene a ser menos probable tener un resultado posible con frecuencia observada igual a cero, por tanto, esperamos un incremento del $\chi^2$ calculado. 

%%Por consiguiente, el conjunto de resultados con cantidades uno o cero son considerados por tanto como muestras \textit{extremas}.

Llamamos \textit{muestra extrema} a una muestra que no ha alcanzado total representatividad, es decir, al menos una combinación posible ha tenido frecuencia observada igual a cero.

\begin{table}[ht]
%%\caption{Cálculo del tamaño muestral requerido para una total representatividad mediante prueba de bondad de ajuste $\chi^2$ con muestras extremas.}
\caption{Cálculo de la bondad de ajuste $\chi^2$ obtenida con un tamaño muestral de $100$ bajo muestra extrema.}
\centering 
\begin{tabular}{c c c c} 
\hline\hline\\[-2ex]
%\multicolumn{4}{c}{Estadístico Chi-Cuadrado con $\alpha=0.05$}\\
%\hline \\[-2ex]
Resultado Posible & Frecuencia Observada & Frecuencia Esperada & $\chi^2$\\ [0.5ex] 
\hline\\[-1ex] 
000 & 14 & 12.5 & 0.18\\ 
001 & 14 & 12.5 & 0.18\\ 
010 & 14 & 12.5 & 0.18\\ 
011 & 15 & 12.5 & 0.50\\  
100 & 15 & 12.5 & 0.50\\ 
101 & 0 & 12.5 & 12.5\\ 
110 & 14 & 12.5 & 0.18\\ 
111 & 14 & 12.5 & 0.18\\ [1ex] % [1ex] adds vertical space
%\hline \\[-1ex]
$Total$ & $100$ & $100$ & $14.40$\\
\hline\hline\\[-2ex]
%\hline \\[-2ex]
\multicolumn{3}{l}{Valor crítico $\chi^2_{(0.05,7)}$} & 14.06714\\
\multicolumn{3}{l}{Valor calculado $\chi^2$} & 14.40000\\
%\hline \\[-2ex]
%\multicolumn{3}{c}{$p$-valor} & 0.04451\\
\hline\hline %inserts single line
\end{tabular}
\label{tab:chisq} % is used to refer this table in the text
\end{table}

La tabla precedente ilustra que el valor calculado experimental de $\chi^2$ supera al valor crítico de 14.06714 cuando el tamaño de la muestra extrema es 100. Si utilizamos esta misma formulación de tabla pero con muestras menores, el valor de $\chi^2$ no alcanza al valor crítico, y si utilizamos la misma formulación con muestras mayores, el valor $\chi^2$ siempre supera al valor crítico.

Consecuentemente la probabilidad de una muestra tan extrema como la del ejemplo, se torna menor que $\alpha$ cuando $m \ge 100$.

\begin{lema*}
Considere una muestra de tamaño $m$ con $k < m$ valores diferentes en el espacio de resultados posibles, suponga además una prueba de bondad de ajuste $\chi^2$ aplicada a la hipótesis nula de que el conjunto de valores observados proceden de una densidad multinomial, utilizando un nivel $\alpha$ de significancia. Entonces, existe un entero $m* > 1$ tal que $\chi^2_{calculado} > \chi^2_{cr\acute{i}tico}$ para cualquier muestra extrema de tamaño $m > m*$.
\end{lema*}

{\it Esbozo de la demostración}: Dada una muestra de tamaño $m$, para una densidad multinomial la cantidad esperada en el resultado posible $i$ es $mp_i$. Por tanto, según el Teorema del Límite Central, esta cantidad resultante se aproxima a su $mp_i$ respectivo, y por ende la probabilidad de obtener una cantidad igual a cero en el resultado posible $i$ decrece monótonamente a medida que $m$ crece.

Por lo tanto, podemos tomar esto como un tamaño muestral recomendado: el $m*$ que produce el primer $\chi^2_{calculado}$ mayor que el $\chi^2_{cr\acute{i}tico}$. De esta forma, tenemos una razonable garantía que el resultado no será extremo, esto es, al menos con probabilidad $1 - \alpha$ ninguna de las frecuencias observadas en los resultados será igual a cero. Para este ejemplo hemos obtenido $m*$ = 100 con $\alpha$ = 0.05.  

\begin{figure}[htb]
	\centering
	\includegraphics[width=.8\columnwidth]{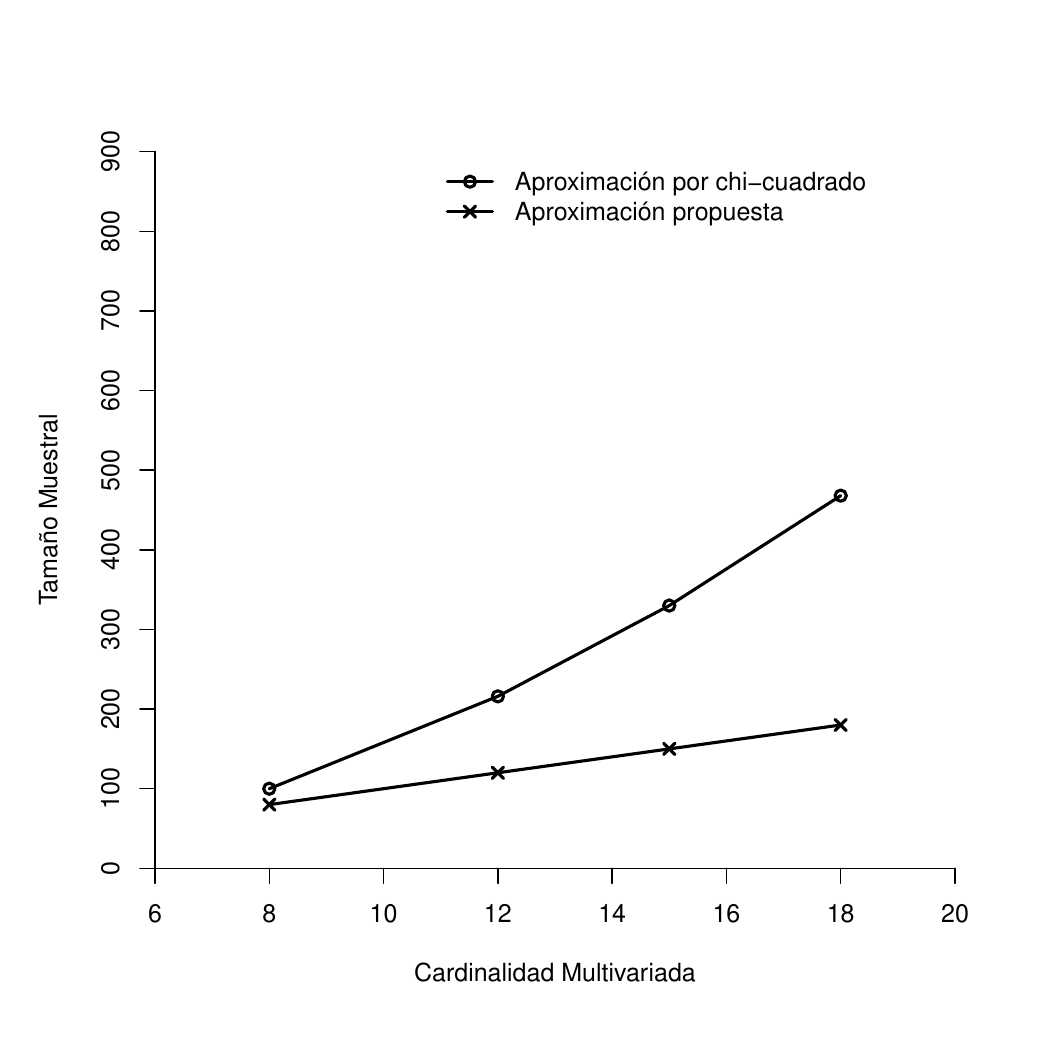}
	\caption{La aproximación del tamaño muestral propuesta y el tamaño muestral recomendado empleando la prueba de bondad de ajuste $\chi^2$ basada en el primer $\chi^2_{calculado} > \chi^2_{cr\acute{i}tico}$ con fuente multinomial equiprobable como $H_{0}$.}
	\label{fig:test_chi}
\end{figure}

Corridas adicionales para combinaciones con $12$, $15$ y $18$ valores equiprobables, han generado tamaños muestrales con $216$, $330$ y $468$ respectivamente bajo la misma significancia $\alpha$. Los cuatro puntos así obtenidos forman casi una línea recta en comparación a la aproximación propuesta como se puede ver en la Figura \ref{fig:test_chi}.

Ahora nos referimos al caso de un espacio de resultados con probabilidades no equitativas. Las distribuciones multinomiales con probabilidades no equitativas alcanzan bajos $mp_i$ como valores esperados en resultados posibles con baja probabilidad, es decir, una cantidad igual a cero en estos resultados posibles es más probable, especialmente para pequeños tamaños muestrales. 

Por otra parte, altos valores de $p_i$ generan altos valores esperados, tal que una cantidad igual a cero en estos resultados posibles es extremadamente improbable. 

En conjunto, los experimentos muestrales demuestran en estos casos que el tamaño muestral requerido es frecuentemente mayor que en el caso de probabilidades equitativas.

No implica dificultad establecer verificaciones similares para un conjunto de atributos categóricos o discretos con cardinalidades univariadas $c_1$, ..., $c_n$. Este procedimiento experimental nos lleva a una aproximación más conservadora, es decir se requerirá mayor tamaño muestral para una razonable garantía de total representatividad.

\section{Resumen}
En este capítulo se ha descrito el escenario experimental y los resultados obtenidos a partir del análisis de sesgo realizado sobre la medida $MSU$ aplicado a variables sintéticas generadas para emular a atributos y clases en una tarea de clasificación.

El estudio del sesgo se ha realizado sobre la mayor diversidad de atributos posibles, esto es, atributos individualmente informativos, no-informativos y colectivamente informativos.

Se ha identificado los factores asociados al sesgo en las mediciones mediante el $MSU$ y se ha propuesto una relación de asociación empírica entre los mismos con el fin de controlar los efectos de dichos factores.

Se ha llevado a cabo una simulación de una prueba de bondad de ajuste $\chi^2$ con el fin de obtener una aproximación de referencia desde la perspectiva de la inferencia estadística.

El contraste con la referencia resultante de la mencionada simulación, nos ha permitido constatar un mejor ajuste del cálculo del tamaño muestral propuesto, así como también un comportamiento lineal ante la variación del cardinalidad multivariada.

Finalmente, de manera a tener un panorama generalizado de los experimentos realizados en el presente capítulo, un resumen de las características de las muestras utilizadas en cada experimento se muestra en la Tabla ~\ref{tabla:resultados_analisis}.

En dicha tabla, la primera columna se refiere a la figura resultante del experimento, luego en la segunda columna, se muestra la regla de clasificación utilizada (Método de Kononenko ó Función Ó-Exclusivo) para determinar la naturaleza informativa de los atributos.

En la tercera columna, se describe el tamaño muestral utilizado con una cantidad fija o variable de instancias. 

Las demás columnas se refieren a la cantidad de atributos informativos, el número de valores posibles, la cantidad de atributos no-informativos y el número de clases respectivamente.

\begin{table}[ht]
\caption{Resumen de las características de cada experimento del capítulo.}
\centering % used for centering table
\begin{tabular}{c c c c c c c} % centered columns (4 columns)
\hline\hline\\[-2ex] %inserts double horizontal lines
Figura & Regla & \#$TM$ & \#$a_i$ & \#e & $a_{n}$ & \#c\\ [0.5ex] % inserts table
%heading
\hline\\[-1ex] % inserts single horizontal line
\ref{fig:figuras-a1-a2}(a) & MK & 1000 & 1 & [2-40]& 1 & 10\\ 
\ref{fig:figuras-a1-a2}(b) & MK & 1000 & 1 & [2-40]& 1 & 2\\
\ref{fig:figuras-e1-e2}(a) & MK & [8-50] & 2 & 2 & 2 & 2\\
\ref{fig:figuras-e1-e2}(b) & MK & [8-150] & 2 & 2 & 2 & 2\\
\ref{fig:figuras-b1-b2}(a) & XOR & [8-50] & 2 & 2 & 0 & 2\\
\ref{fig:figuras-b1-b2}(b) & XOR & [8-150] & 2 & 2 & 0 & 2\\
\ref{fig:figuras-cd}(a) & MK & 5000 & [4-12] & [4-64] & 0 & 2 \\
\ref{fig:figuras-cd}(b) & - & 5000 & 0 & [4-64] & [4-12] & 2 \\
\ref{fig:test_chi} & - & [0-500] & 0 & 2& [2-4] & 2 \\ [1ex]
\hline\hline %inserts single line
\end{tabular}
\label{tabla:resultados_analisis} % is used to refer this table in the text
\end{table}

  \SetKwInput{Kw}{Entrada} 
\chapter{RESULTADOS}

En el presente capítulo presentaremos los resultados experimentales obtenidos a través de las simulaciones realizadas sobre los datos sintéticos generados y a partir de la aproximación propuesta en el análisis descrito en el capítulo anterior.

\section{Propósito de los experimentos}
El escenario experimental cuyos detalles se explican en el siguiente apartado, ha sido montado con el objetivo de verificar el comportamiento del $MSU$ basado en los factores de sesgo identificados y principalmente en función a la relación de asociación propuesta para controlar dicho sesgo.

Atendiendo la naturaleza informativa (individual o colectiva) y no-informativa de los atributos, los experimentos han sido llevados a cabo en la siguiente diversidad de escenarios posibles:

\begin{itemize}
\item Variación de la cardinalidad univariada para subconjuntos de atributos individualmente informativos y no-informativos.
\item Variación en la cantidad de atributos que conforman subconjuntos de atributos tanto individualmente informativos, como así también no-informativos y colectivamente informativos.
\item Variación de ruido para un conjunto de atributos colectivamente informativos mediante la inclusión de atributos no informativos. 
\item Variación de ruido para un conjunto de atributos colectivamente informativos mediante la inclusión de atributos individualmente informativos. 
\end{itemize}

\section{Escenario Experimental}
Se ha adoptado de la fase de análisis descrito en el capítulo anterior tanto la técnica de simulación Monte Carlo, como así también los datos sintéticos y el entorno computacional.

Al mencionado escenario experimental, se ha agregado un conjunto de experimentos con un tamaño muestral fijo de $600$ instancias y por otra parte, se ha agregado el cálculo del tamaño muestral mediante la aproximación propuesta
\begin{equation}
Tama\tilde{n}o\ muestral \approx 10\left\vert{clase}\right
\vert\prod_{i=1}^{n} \left\vert{f_{i}}\right\vert.
\end{equation}

\section{Descripción de los resultados}
En la presente sección se llevará a cabo una demostración e interpretación de los resultados obtenidos. 

\subsection{Resultado Comparativo \#1}

En este par de experimentos mostrado en la Figura \ref{fig:figuras-f1-f2}, se ha verificado el efecto de la variación de la cardinalidad univariada para dos subconjuntos, uno de ellos conformado por atributos individualmente informativos y el otro por atributos no-informativos.

\subsubsection{Tamaño muestral fijo}
En la Figura \ref{fig:figuras-f1-f2} (a) se muestra el efecto de la variación de la cardinalidad univariada para un tamaño muestral fijo de $5000$ instancias.

Como se puede apreciar, para el subconjunto de atributos individualmente informativos, el $MSU$ tiene un punto de inflexión en su medición específicamente en la cardinalidad $10$ ya que a partir de dicho punto el tamaño muestral es insuficiente conforme la aproximación propuesta.

Por otra parte, para el subconjunto de atributos no-informativos, el $MSU$ presenta un crecimiento mayor en su medición específicamente en la cardinalidad $10$ ya que a partir de dicho punto el tamaño muestral es insuficiente conforme la aproximación propuesta.

\subsubsection{Tamaño muestral calculado}

\begin{figure}[!htb]
    \subfloat[Efectos del tamaño muestral fijo de $5000$ instancias y la cardinalidad univariada sobre subconjuntos de atributos informativos y no-informativos.  
              \label{subfig:f2}]{
      \includegraphics[width=0.5\textwidth]{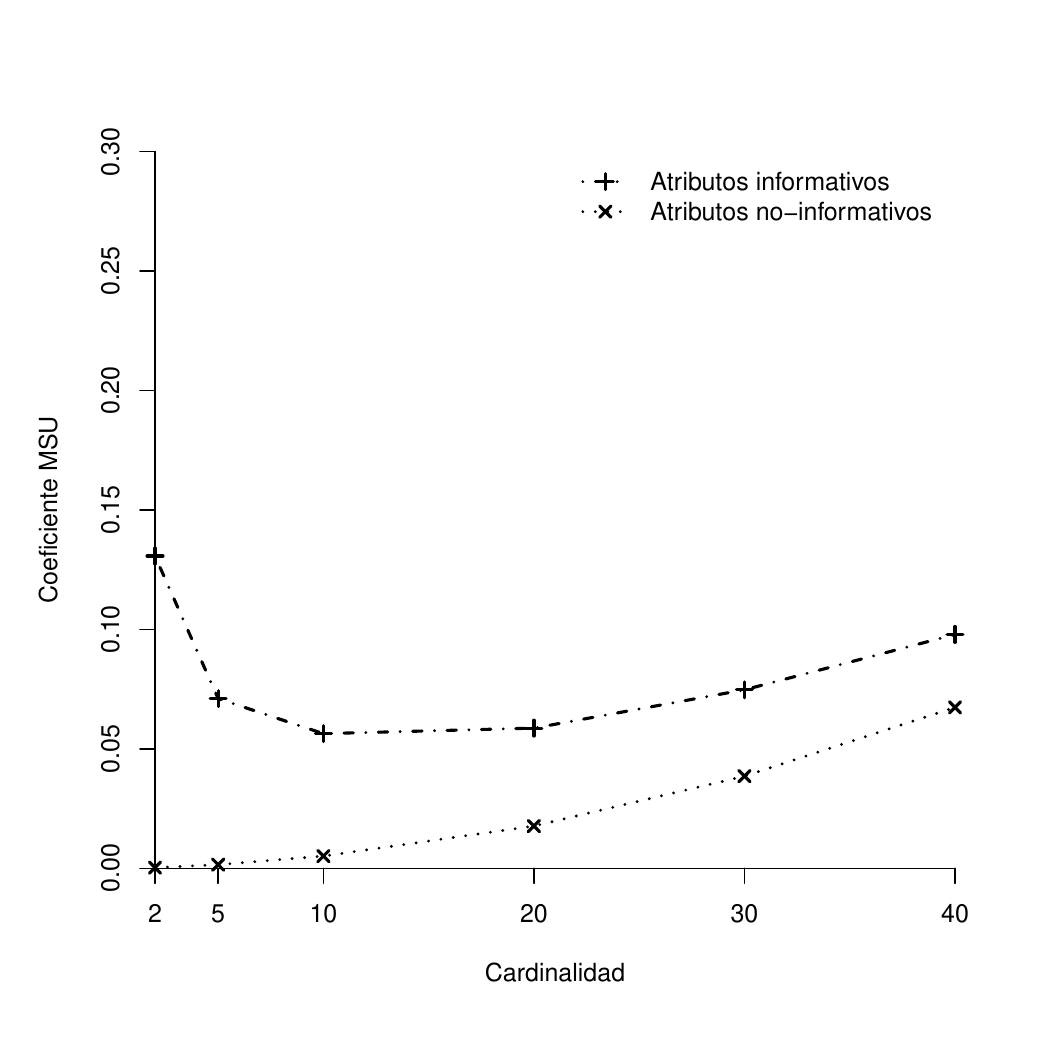}
    }
    \quad
    \subfloat[Efectos del tamaño muestral calculado y la cardinalidad univariada  sobre subconjuntos de atributos informativos y no-informativos.
              \label{subfig:f}]{
      \includegraphics[width=0.5\textwidth]{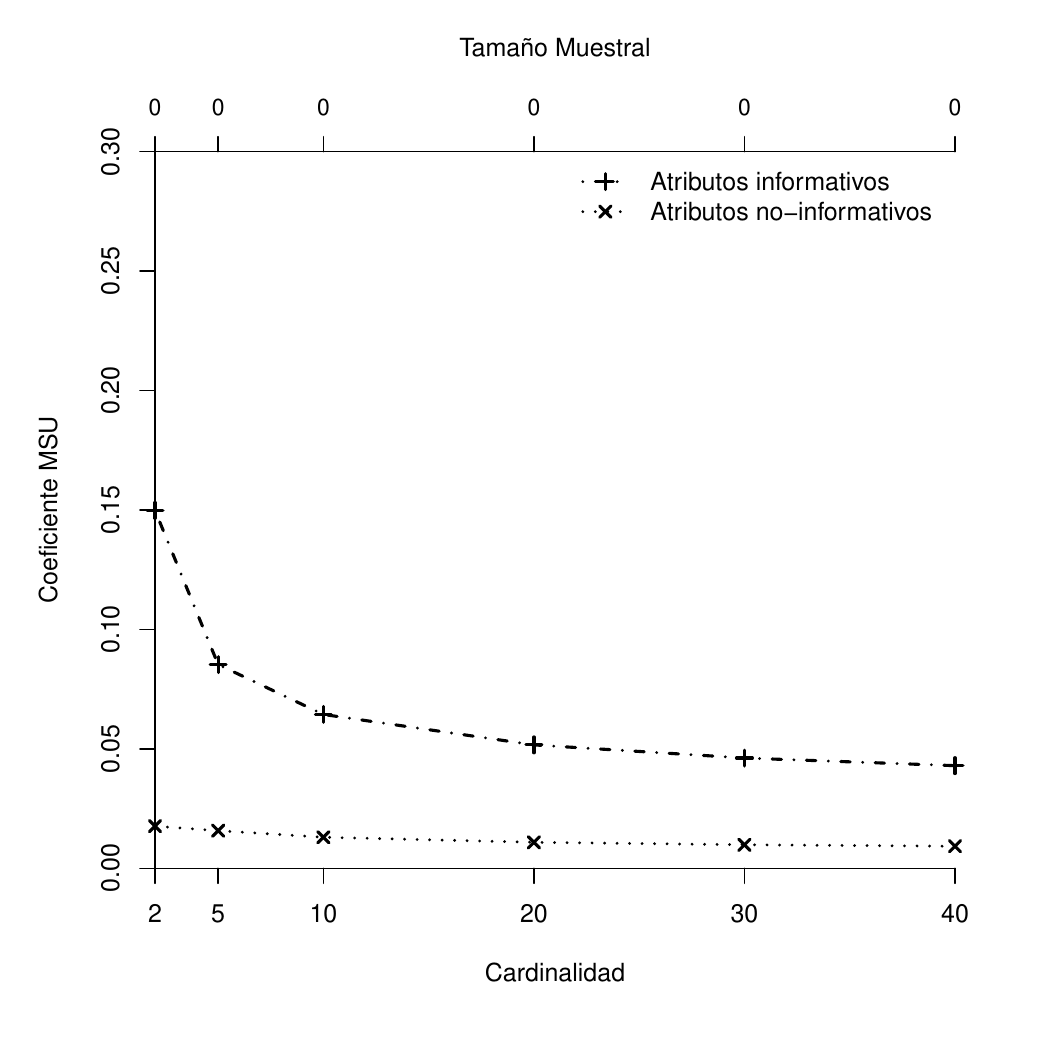}
    }
    \caption{Efectos del tamaño muestral fijo y calculado así como de la cardinalidad univariada de los atributos sobre el MSU. La cardinalidad de la clase es $2$.}
    \label{fig:figuras-f1-f2}
\end{figure}

En la Figura \ref{fig:figuras-f1-f2} (b) se muestra el efecto de la variación de la cardinalidad univariada para un tamaño muestral calculado conforme la aproximación propuesta.

Como se puede apreciar y en contraste al tamaño muestral fijo, para el subconjunto de atributos individualmente informativos, el $MSU$ tiene un suave decrecimiento con tendencia a estabilizarse en punto mayor a cero.

Si bien, este comportamiento no es el esperado, el $MSU$ ya no presenta un sesgo marcado por un punto de inflexión a partir del cual el mismo exhibe cambio inverso en la tendencia informativa del subconjunto.

Por otra parte, para el subconjunto de atributos no-informativos, el $MSU$ presenta una medición estable próximo al cero a medida que se incrementa la cardinalidad univariada de los atributos, lo cual es el comportamiento esperado en contraste al sesgo exhibido con el tamaño muestral fijo.

\subsection{Resultado Comparativo \#2}

En este par de experimentos mostrado en la Figura \ref{fig:figuras-gh}, se ha verificado el efecto de la variación en el número de atributos que conforman un total de tres subconjuntos, el primero de ellos conformado por atributos individualmente informativos, el segundo por atributos no-informativos y el tercero por atributos colectivamente informativos.

\subsubsection{Tamaño muestral fijo}
En la Figura \ref{fig:figuras-gh} (a) se muestra el efecto de la variación del número de atributos que conforman los subconjuntos para un tamaño muestral fijo de $1000$ instancias.

Para el \textit{subconjunto de atributos individualmente informativos}, se puede apreciar que el $MSU$ presenta un valor inicial mayor a cero con un crecimiento suave seguido de uno exponencial en la medición del nivel de información del subconjunto a medida que se van agregando atributos al mismo, lo cual representa un sesgo.

Para el \textit{subconjunto de atributos no-informativos}, se puede apreciar que el $MSU$ presenta un valor inicial igual a cero hasta cierto punto donde el mismo presenta un crecimiento exponencial en la medición del nivel de información del subconjunto a medida que se van agregando atributos al mismo, lo cual representa un sesgo.

Finalmente, para el \textit{subconjunto de atributos colectivamente informativos}, se puede apreciar que el $MSU$ presenta un valor inicial mayor a cero con un decrecimiento exponencial seguido de un crecimiento exponencial en la medición del nivel de información del subconjunto a medida que se van agregando atributos al mismo, lo cual representa un sesgo.

\subsubsection{Tamaño muestral calculado}
\begin{figure}[!htb]
    \subfloat[Efectos del tamaño muestral sobre un conjunto de atributos informativos individualmente, colectivamente informativos y no-informativos. El tamaño muestral es fijo de $1000$ instancias. 
              \label{subfig:g}]{
      \includegraphics[width=0.5\textwidth]{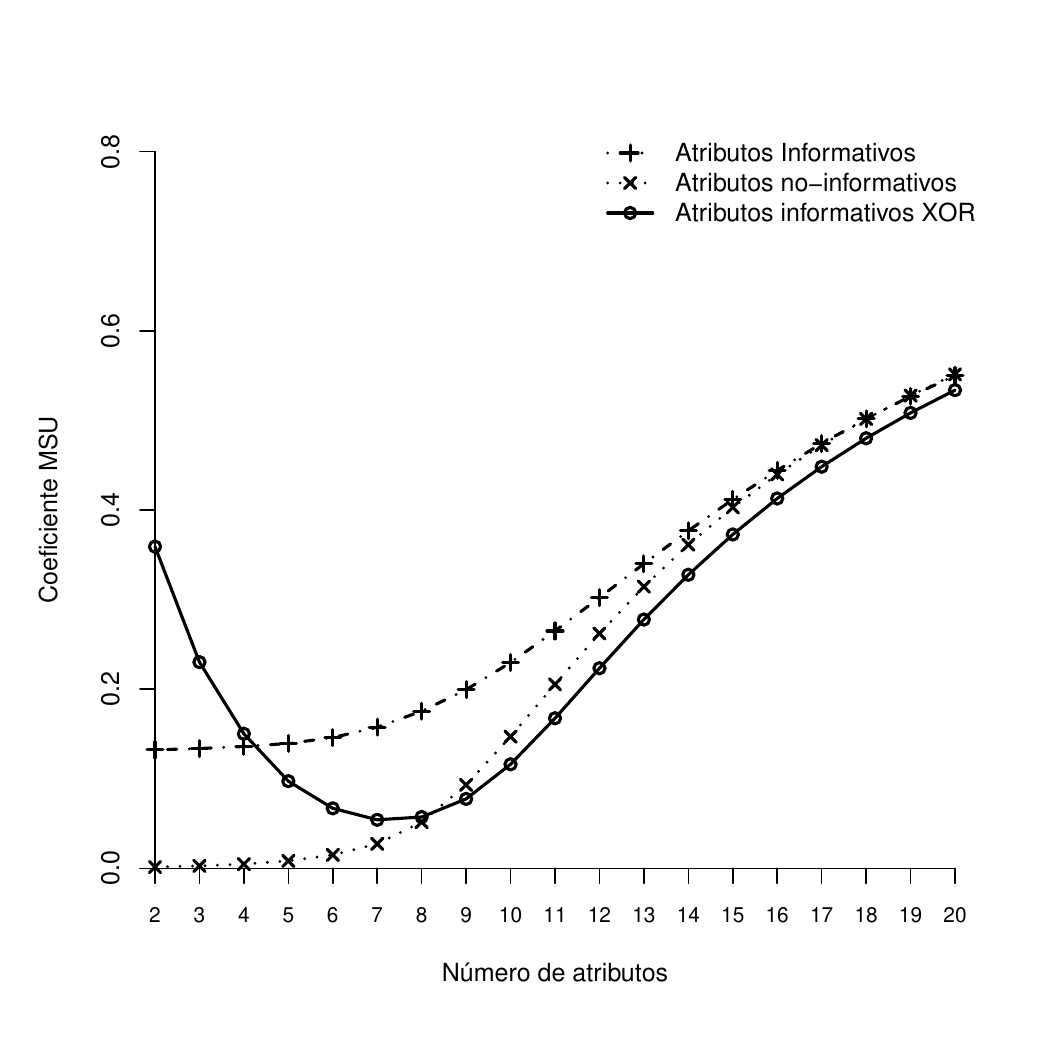}
    }
    \quad
    \subfloat[Efectos del tamaño muestral calculado sobre un conjunto de atributos informativos individualmente, colectivamente informativos y no-informativos.
              \label{subfig:h}]{
      \includegraphics[width=0.5\textwidth]{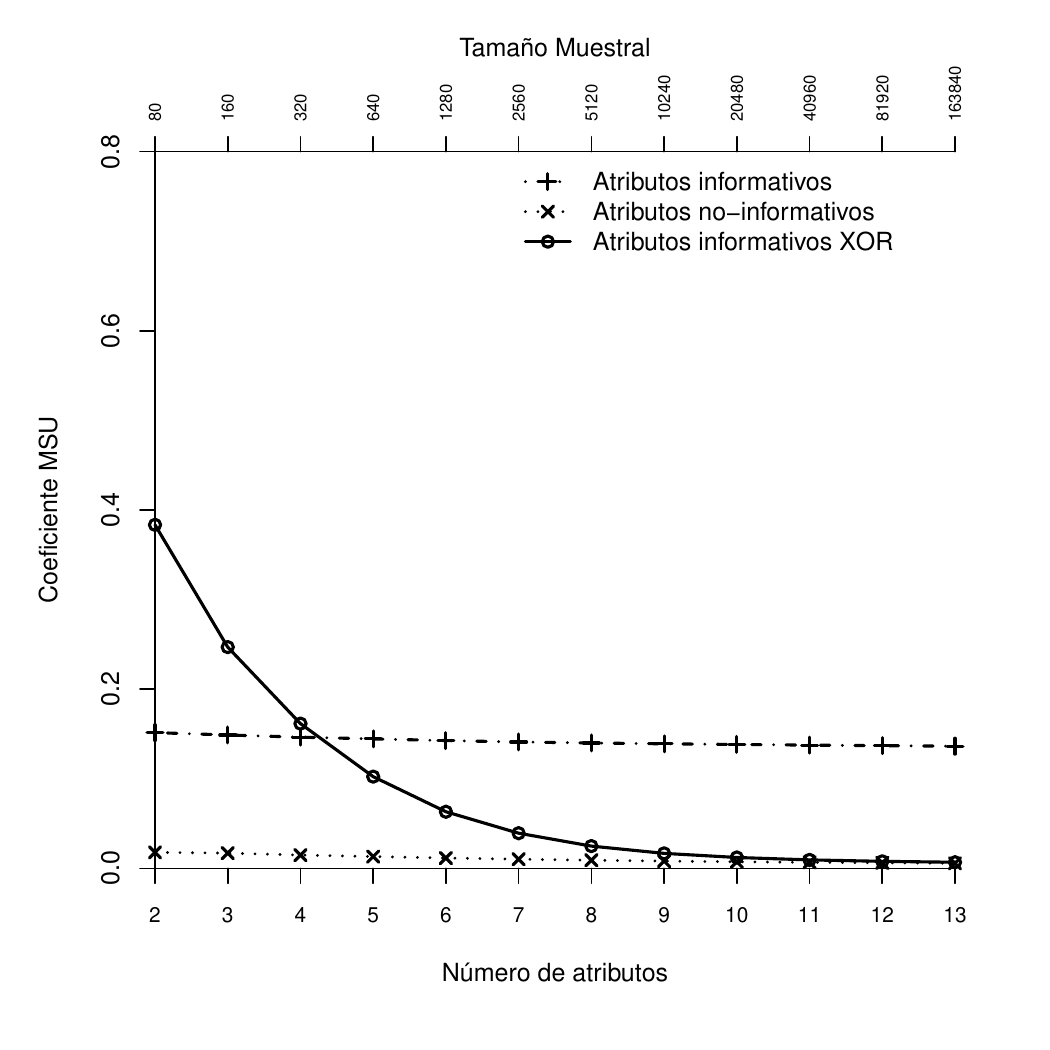}
    }
    \caption{Los efectos de la variación del número de atributos con tamaño muestral fijo y calculado  sobre el MSU. La cardinalidad univariada es $2$.}
    \label{fig:figuras-gh}
\end{figure}

En la Figura \ref{fig:figuras-gh} (a) se muestra el efecto de la variación del número de atributos que conforman los subconjuntos para un tamaño calculado conforme la aproximación propuesta.

Para el \textit{subconjunto de atributos individualmente informativos}, se puede apreciar que el $MSU$ presenta un comportamiento estable con un valor mayor a cero en la medición del nivel de información del subconjunto a medida que se van agregando atributos al mismo, lo cual representa el comportamiento esperado con el sesgo bajo control.

Para el \textit{subconjunto de atributos no-informativos}, se puede apreciar que el $MSU$ presenta un comportamiento estable con un valor muy próximo a cero en la medición del nivel de información del subconjunto a medida que se van agregando atributos al mismo, lo cual representa el comportamiento esperado con el sesgo controlado.

Finalmente, para el \textit{subconjunto de atributos colectivamente informativos}, se puede apreciar que el $MSU$ presenta un valor inicial mayor a cero con un suave decrecimiento que tiende a cero en la medición del nivel de información del subconjunto a medida que se van agregando atributos al mismo.

Este comportamiento es el esperado, ya que al agregar atributos al subconjunto, simplemente la probabilidad que la combinación de sus valores represente un $XOR$ va tendiendo a cero y por ende su nivel de información colectiva. Por ejemplo, si se tiene $n$ atributos, uno de ellos deberá poseer un valor $Verdadero$ y el resto ó $n-1$ indefectiblemente un valor $Falso$, esto es, de tal forma a que toda la instancia represente un $XOR$.

\subsection{Resultado Comparativo \#3}
En este par de experimentos mostrado en la Figura \ref{fig:figuras-xor-1_2}, se ha verificado el efecto de agregar ruido a un conjunto de atributos colectivamente informativos mediante la inclusión de atributos no-informativos.

\subsubsection{Tamaño muestral fijo}
En la Figura \ref{fig:figuras-xor-1_2} (a) se muestra el efecto de la inclusión de atributos no-informativos a un conjunto de atributos colectivamente informativos para un tamaño muestral fijo de $600$ instancias.

Como se puede apreciar claramente, a medida que se va agregando atributos no-informativos el $MSU$ ó nivel de información del conjunto decrece hasta cierto punto, lo cual es lo esperado. Sin embargo, a partir de dicho punto el $MSU$ tiene un crecimiento exponencial representando un sesgo en la medición del nivel de información del conjunto.

\subsubsection{Tamaño muestral calculado}

\begin{figure}[!htb]
    \subfloat[Efectos de agregar atributos no-informativos a un par de atributos XOR. El tamaño muestral es fijo de $600$ instancias.
              \label{subfig:xor_1}]{
      \includegraphics[width=0.5\textwidth]{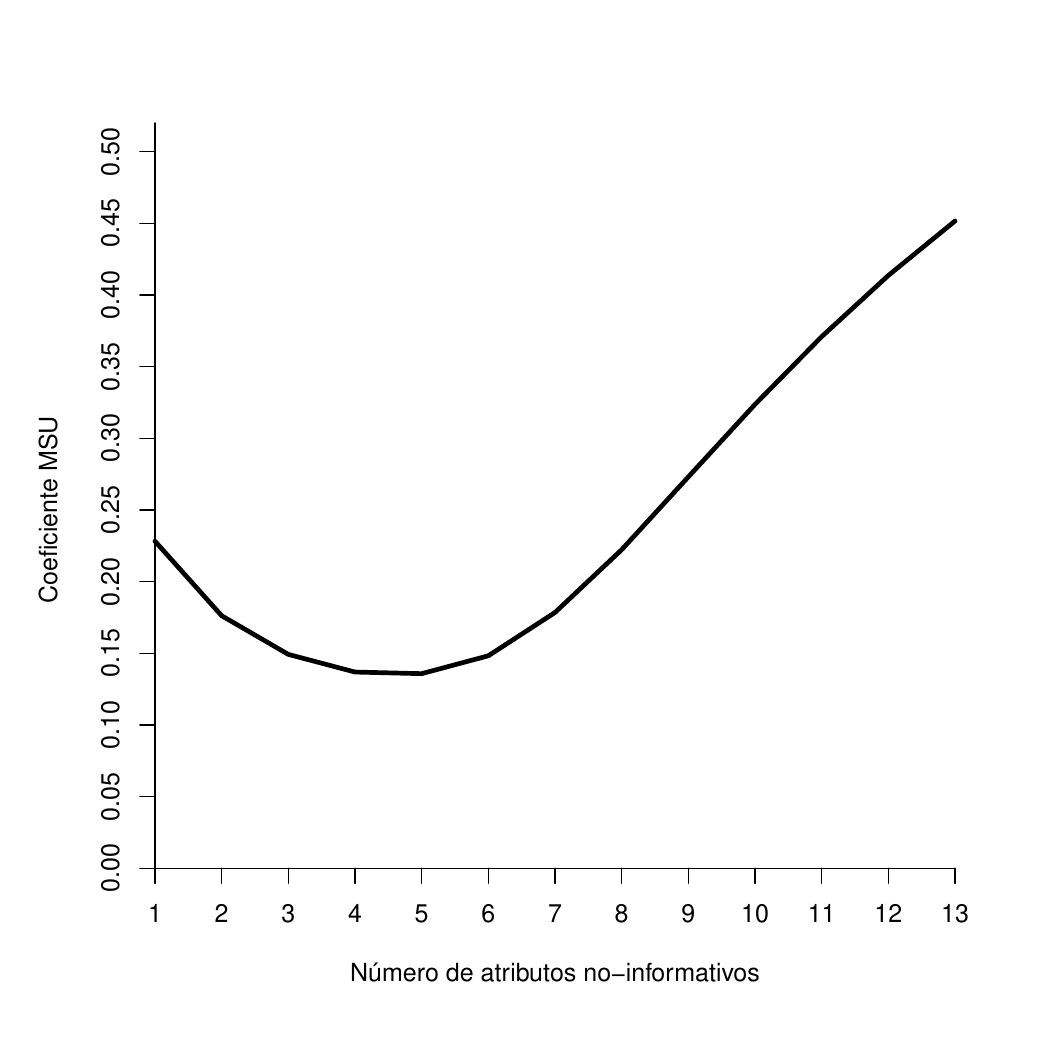}
    }
    \quad
    \subfloat[Efectos de agregar atributos no-informativos a un par de atributos XOR. El tamaño muestral es calculado en función a la aproximación propuesta.
              \label{subfig:xor_2}]{
      \includegraphics[width=0.5\textwidth]{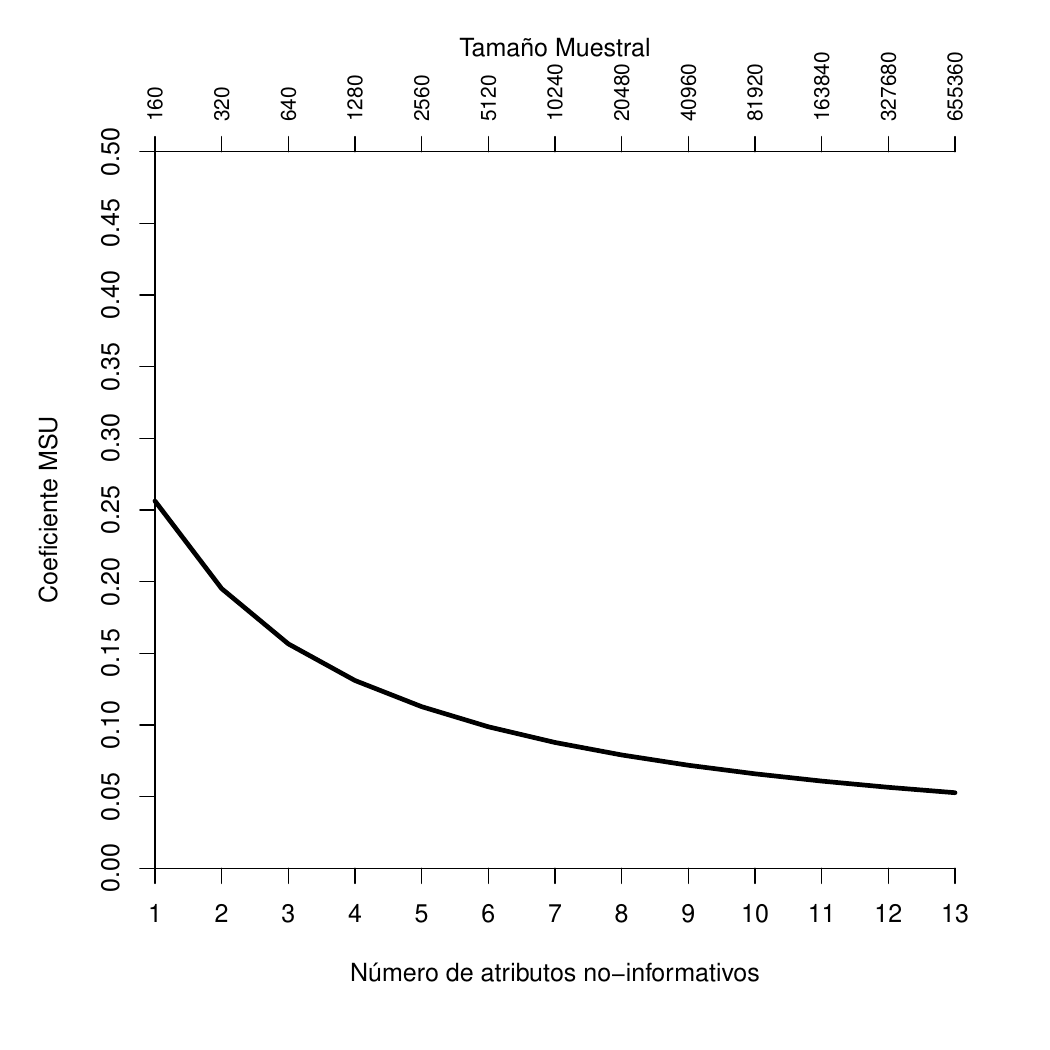}
    }
    \caption{Los efectos de agregar atributos no-informativos con tamaño muestral fijo y calculado sobre el MSU. La cardinalidad univariada es $2$.}
    \label{fig:figuras-xor-1_2}
\end{figure}

En la Figura \ref{fig:figuras-xor-1_2} (b) se muestra el efecto de la inclusión de atributos no-informativos a un conjunto de atributos colectivamente informativos para un tamaño muestral calculado conforme la aproximación propuesta.

Como se puede apreciar y en contraste al tamaño muestral fijo, a medida que se va agregando atributos no-informativos, el $MSU$ del conjunto presenta un decrecimiento que tiende a cero en el nivel de información del mismo, lo cual es el comportamiento esperado debido un sesgo bajo control.

\subsection{Resultado Comparativo \#4}
En este par de experimentos mostrado en la Figura \ref{fig:figuras-xor-3_4}, se ha verificado el efecto de agregar ruido a un conjunto de atributos colectivamente informativos mediante la inclusión de atributos individualmente informativos.

\subsubsection{Tamaño muestral fijo}

En la Figura \ref{fig:figuras-xor-3_4} (a) se muestra el efecto de la inclusión de atributos individualmente informativos a un conjunto de atributos colectivamente informativos para un tamaño muestral fijo de $600$ instancias.

Como se puede apreciar claramente, a medida que se va agregando atributos no-informativos el $MSU$ ó nivel de información del conjunto decrece suavemente hasta cierto punto, lo cual es lo esperado. Sin embargo, a partir de dicho punto el $MSU$ tiene un suave crecimiento representando un sesgo en la medición del nivel de información del conjunto.

\subsubsection{Tamaño muestral calculado}

\begin{figure}[!htb]
    \subfloat[Efectos de agregar atributos individualmente informativos a un par de atributos XOR. El tamaño muestral es fijo de $600$ instancias.
              \label{subfig:xor_3}]{
      \includegraphics[width=0.5\textwidth]{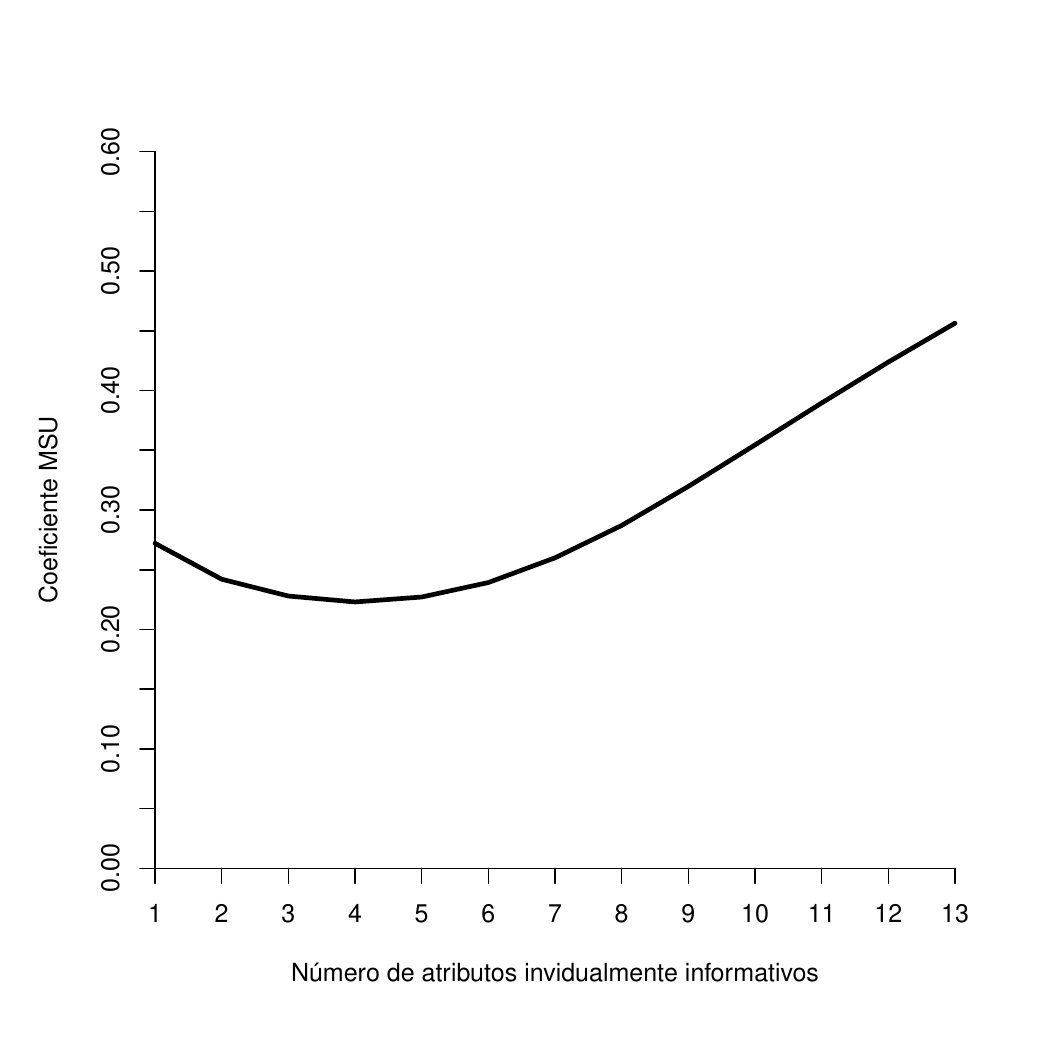}
    }
    \quad
    \subfloat[Efectos de agregar atributos individualmente informativos a un par de atributos XOR. El tamaño muestral es calculado en función a la aproximación propuesta.
              \label{subfig:xor_4}]{
      \includegraphics[width=0.5\textwidth]{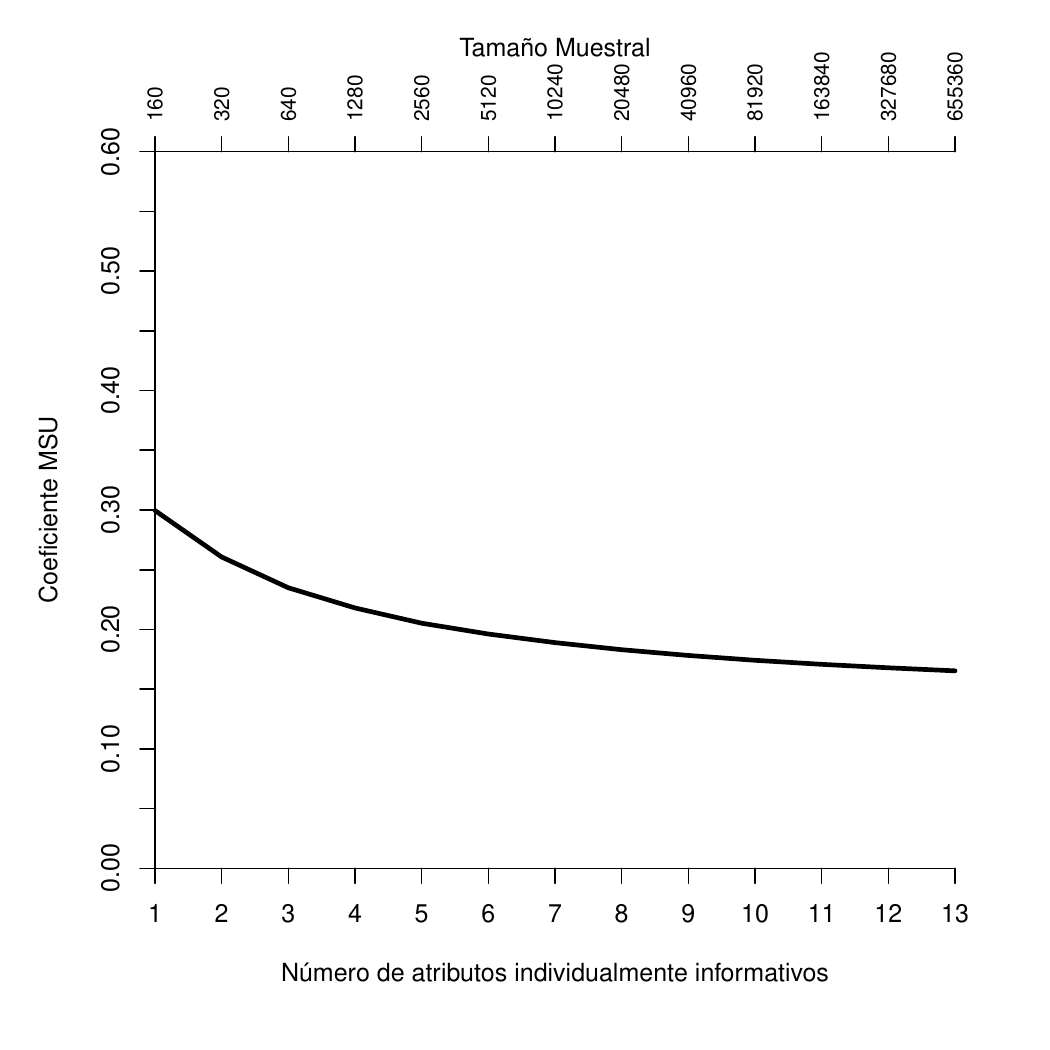}
    }
    \caption{Los efectos de atributos individualmente informativos con tamaño muestral fijo y calculado  sobre el MSU. La cardinalidad univariada es $2$.}
    \label{fig:figuras-xor-3_4}
\end{figure}

En la Figura \ref{fig:figuras-xor-3_4} (b) se muestra el efecto de la inclusión de atributos individualmente informativos a un conjunto de atributos colectivamente informativos para un tamaño muestral calculado conforme la aproximación propuesta.

Como se puede apreciar y en contraste al tamaño muestral fijo, a medida que se va agregando atributos individualmente informativos, el $MSU$ del conjunto presenta un suave decrecimiento que tiende a estabilizarse en un valor mayor a cero en el nivel de información del mismo.

Este comportamiento es el esperado, ya que si bien se van agregando atributos informativos, los mismos poseen información acerca de la clase de forma individual y al ser incorporados al conjunto a ser evaluado por el $MSU$ van degradando lentamente (en comparación a atributos no-informativos) el nivel de información del mismo en conjunción.

\section{Resumen}
En este capítulo se ha descrito el escenario experimental, los experimentos y la interpretación de los resultados obtenidos, esto es, a partir de la aplicación de la expresión analítica propuesta en esta tesis, con el fin de controlar el sesgo sobre la medida $MSU$ sobre variables sintéticas generadas para emular a los atributos y la clase en una tarea de clasificación.

El estudio del control del sesgo mediante la asociación propuesta, se ha realizado sobre la mayor diversidad de atributos posibles, esto es, atributos individualmente informativos, no-informativos y colectivamente informativos.

Los resultados demuestran que mediante la expresión analítica propuesta se logra un comportamiento controlado en cuanto al sesgo en las mediciones mediante el $MSU$, lo cual abre un abanico de oportunidades experimentales en el proceso de selección de atributos.

De manera a tener un panorama generalizado de los experimentos realizados en el presente capítulo, un resumen de las características de las muestras utilizadas en cada experimento se muestra en la Tabla ~\ref{tabla:resumen_resultados}.

En dicha tabla, la primera columna se refiere a la figura resultante del experimento, luego en la segunda columna, se muestra la regla de clasificación utilizada (Método de Kononenko ó Función Ó-Exclusivo) para determinar la naturaleza informativa de los atributos.

En la tercera columna, se describe el tamaño muestral utilizado ya sea con una cantidad fija de instancias o de lo contrario como consecuencia del cálculo realizado mediante la expresión analítica propuesta, el cual lo hemos simbolizado como $g$. 

Las demás columnas se refieren a la cantidad de atributos informativos, el número de valores posibles, la cantidad de atributos no-informativos y el número de clases respectivamente.

\begin{table}[ht]
\caption{Resumen de las características de cada experimento donde $g$ se refiere a la expresión analítica propuesta para estimar el tamaño muestral $(TM)$.}
\centering % used for centering table
\begin{tabular}{c c c c c c c} % centered columns (4 columns)
\hline\hline\\[-2ex] %inserts double horizontal lines
Figura & Regla & \#$TM$ & \#$a_i$ & \#e & $a_{n}$ & \#c\\ [0.5ex] % inserts table
%heading
\hline\\[-1ex] % inserts single horizontal line
\ref{fig:figuras-f1-f2}(a) & MK & 5000 & 2 & [2-40]& 2 & 2\\ 
\ref{fig:figuras-f1-f2}(b) & MK & $g$  & 2 & [2-40]& 2 & 2\\
\ref{fig:figuras-gh}(a) & ambos & 1000 & [2-20] & 2 & [2-13] & 2\\
\ref{fig:figuras-gh}(b) & ambos & $g$  & [2-20] & 2 & [2-13] & 2\\
\ref{fig:figuras-xor-1_2}(a) & XOR & 600 & 2 & 2 & [1-13] & 2\\
\ref{fig:figuras-xor-1_2}(b) & XOR & $g$ & 2 & 2 & [1-13] & 2\\
\ref{fig:figuras-xor-3_4}(a) & ambos & 600 & [3-15] & 2 & 0 & 2 \\
\ref{fig:figuras-xor-3_4}(b) & ambos & $g$ & [3-15] & 2 & 0 & 2 \\ [1ex] % [1ex] adds vertical space
\hline\hline %inserts single line
\end{tabular}
\label{tabla:resumen_resultados} % is used to refer this table in the text
\end{table}

  \SetKwInput{Kw}{Entrada} 
\chapter{CONCLUSIONES}

Dada su naturaleza de medida de correlación de variables agrupadas, en este trabajo, se ha caracterizado el comportamiento del $MSU$ mediante un análisis de sesgo en el contexto del proceso de selección de atributos.

Se ha demostrado que los factores implicados en el sesgo del $MSU$ para la correspondiente detección de la interacción de variables agrupadas en conjunto son el número de variables, la cardinalidad univariada, la cardinalidad multivariada y el tamaño muestral.

Dado el análisis individual de los mencionados factores de sesgo en el $MSU$, podemos concluir que la cardinalidad multivariada es el factor que engloba a los demás factores ya que el mismo se calcula en base a las cardinalidades univariadas y por otra parte, se ha demostrado igualmente, que el tamaño muestral debe poder albergar al menos a todas las combinaciones de valores posibles de las variables del conjunto.

%%En efecto, y haciendo una analogía, la cardinalidad multivariada es el número de individuos diferentes que integran una población y por lo tanto, se ha propuesto que el tamaño muestral debe estar dado en función a un múltiplo de veces la cardinalidad multivariada, esto es, de tal forma a garantizar que exista la probabilidad de \textit{total representatividad} en la muestra.  

Por tanto, proponemos que el tamaño muestral esté dado en función a un múltiplo de la cardinalidad multivariada, esto es, de tal forma a facilitar una alta probabilidad de \textit{total representatividad} en la muestra.  

En este sentido, los resultados comparativos obtenidos mediante los experimentos realizados, reflejan una clara tendencia para controlar el sesgo en el $MSU$ dada la relación de asociación que se ha propuesto en el presente trabajo

\begin{equation}
Tama\tilde{n}o\ muestral \approx 10\left\vert{clase}\right
\vert\prod_{i=1}^{n} \left\vert{f_{i}}\right\vert.
\end{equation}

Puesto que dado un conjunto de datos, los valores de los mencionados factores son conocidos a \textit{priori}, la relación de asociación empírica propuesta permite establecer criterios para la conformación de subconjuntos a ser evaluados por el $MSU$ como parte del proceso de selección de atributos. 

%Nosotros hemos establecido que los factores implicados 
%con el sesgo en la deteccion de interacciones entre 
%diferentes features son la cardinalidad univariada, 
%la cardinalidad multivariada y el sample size.
%--
%Puesto que dado un dataset, los valores de estos 
%factores son conocidos a $priori$, nosotros proponemos 
%una relacion de asociacion empirica entre estos factores 
%que permiten establecer criterios para la conformacion 
%de subsets a ser evaluados por el MSU como parte del 
%proceso de feature selection.
%--
%En el campo del feature selection el estudio de considerar posibles 
%interacciones que pueda existir entre diferentes features cobra 
%cada dia mas relevancia dado que con el rapido avance de la 
%tecnologia nos enfrentamos a espacios con dimensionalidad 
%cada vez mas alta.
%--
%Dado que el sesgo en la medicion de informacion de un atributo es 
%un importante problema, el proposito de esta investigacion fue 
%analizar el rendimiento de la medida MSU sobre diferentes tipos y 
%combinaciones de atributos.

\section{Principales contribuciones}
Las principales contribuciones de este trabajo son:
\begin{itemize}
\item El estudio de sesgo sobre una medida multivariada como lo es el $MSU$.
\item La identificación de los factores implicados en el comportamiento del $MSU$ como medida de correlación de variables agrupadas.
\item La introducción del concepto de total representatividad de la muestra como principio explicativo del comportamiento del $MSU$.
\item Una relación de asociación empírica entre los factores determinados que permite un comportamiento controlado del $MSU$ y que a su vez proporciona criterios para la conformación de subconjuntos a ser evaluados por el $MSU$ como parte del proceso de selección de atributos.
\item Un análisis de la relación de asociación empírica propuesta desde el punto de vista de la inferencia estadística mediante la simulación de pruebas de bondad de ajuste. Este análisis permite hallar el tamaño de muestra $m$ requerido para un conjunto dado de atributos.
\end{itemize}

\section{Trabajos futuros}
Posibles trabajos futuros son:
\begin{itemize}
\item El análisis del $MSU$ en conjuntos de datos reales de alta dimensionalidad en varios dominios.
\item Pruebas de desempeño del $MSU$ bajo densidades de datos conocidas. 
\item El estudio del factor multiplicativo constante de la aproximación propuesta para un mejor ajuste. 
\item La implementación del análisis de sesgo como una fase componente del proceso de selección de atributos y que permita la conformación mesurada de subconjuntos de variables como un pre-procesamiento para la consecución de un escenario con sesgo controlado.
\item El ensayo de la presente metodología para el análisis de otras medidas de correlación de variables categóricas.
\end{itemize}

  \backmatter
  \bibliographystyle{alpha}
%  \bibliography{bibliografia}

\newcommand{\etalchar}[1]{$^{#1}$}

%  \appendix
%  \include{appenA}

\end{document}